\documentclass{aa} 

\usepackage{natbib}
\usepackage{graphicx}
\usepackage{amsmath} 
\usepackage{ulem}
\usepackage{amstext}
\usepackage{color} 
\usepackage{longtable}
\usepackage{lscape}
\usepackage{diagbox}
\usepackage{subfigure}

\usepackage{hyperref}
\usepackage{txfonts}
\definecolor{mulberry}{rgb}{0.77, 0.29, 0.55}
\newcommand{\dm}[1]{{\color{mulberry} #1}}

\begin{document} 

%


\title{Activity time series of old stars from late F to early K\\ VI. Exoplanet mass characterisation and detectability in radial velocity }

\titlerunning{Exoplanet mass characterisation and  detectability in radial velocity}

\author{N. Meunier \inst{1}, R. Pous \inst{1}, S. Sulis \inst{2}, D. Mary\inst{3}, A.-M. Lagrange \inst{4,1}
  }
\authorrunning{Meunier et al.}

\institute{
Univ. Grenoble Alpes, CNRS, IPAG, F-38000 Grenoble, France \email{nadege.meunier@univ-grenoble-alpes.fr}
\and 
Universit\'e Aix Marseille, CNRS, CNES, LAM, Marseille, France
\and 
Universit\'e C\^ote d'Azur, Observatoire de la C\^ote d'Azur, CNRS, Lagrange UMR 7293, CS 34229, 
 06304, Nice Cedex 4, France
\and
LESIA (UMR 8109), Observatoire de Paris, PSL Research University, CNRS, UMPC, Univ. Paris Diderot, 5 Place Jules Janssen, 92195 Meudon, France \\
     }

\offprints{N. Meunier}

\date{Received 22 February 2023 ; Accepted 5 June 2023}

\abstract{Stellar variability impacts radial velocities (hereafter RVs) at various timescales and therefore the detectability of exoplanets and the mass determination based on this technique. Detecting and characterising Earth-like planets in the habitable zone of solar-type stars represents an important challenge in the coming years, however. It is therefore necessary to implement systematic studies  of this issue, for example to delineate the current limitations of RV techniques. } 
 {A first aim of this paper is to investigate  whether the targeted 10\% mass uncertainty from RV follow-up of transits detected by PLATO can be  reached. A second aim of this paper is to analyse and quantify   Earth-like planet detectability for various spectral types.   }
{For this purpose, we implemented blind tests based on a  large data set (more than 20000) of realistic synthetic time series reproducing different phenomena leading to stellar variability such as magnetic activity patterns similar to the solar configuration as well as flows (oscillations, granulation, and supergranulation), covering F6-K4 stars and a wide range of activity levels.   }
{
We find that the 10\% mass uncertainty for a 1~M$_{\rm Earth}$ in the habitable zone of a G2 star cannot be reached, even with an improved version of the usual  correction of stellar activity (here based on a non-linear relation with $\log R'_{HK}$ and cycle phase instead of a linear correlation) and even for long-duration (ten years) well-sampled observations. This level can be reached, however, for masses above 3~M$_{\rm Earth}$ or for K4 stars alone. We quantify the  maximum dispersion of the RV residuals needed to reach this 10\% level, assuming the activity correction method and models do not affect the planetary signal. 
Several other methods, also based on a correction using $\log R'_{HK}$ in various ways (including several denoising techniques and Gaussian processes) or photometry,
were  tested and do not allow a significantly improvement of this limited performance. Similarly,  such low-mass planets in the habitable zone cannot be detected with a similar correction: blind tests lead to  very low detection rates for 1~M$_{\rm Earth}$ and to a very high level of 
false positives. 
We also studied the residuals after correction of the stellar signal, and found significant power in the periodogram at short and long timescales, corresponding to masses higher than 1~M$_{\rm Earth}$ in this period range. 
}
{We conclude that very significant and new improvements  with respect to methods based on activity indicators to correct for stellar activity must be devised at all timescales to reach the objective of 10\% uncertainty on the mass or to detect such planets in RV. Methods based on the correlation with activity indicators are unlikely to be sufficient.  }


\maketitle


\section{Introduction}

Stellar variability has long been recognised to impact radial velocity (RV) measurements and therefore exoplanet detection and characterisation based on this technique 
\cite[][]{saar97,desort07}. It was later shown that for a star such as the Sun, the convective blueshift  dominates the signal \cite[][]{lagrange10b,meunier10a,meunier10} by typically two orders of magnitude compared to the Earth RV signal. In addition, the RV technique, unlike astrometry for example \cite[][]{makarov10b,lagrange11,meunier20,meunier22b}, is impacted by many processes related to flows at different  scales  \cite[see the review in][]{meunier21,meunier22c}.

To characterise the impact of stellar variability, several types of blind tests have been implemented. The main test was organised by \cite{dumusque16}, with  several teams contributing to analyse a small set of time series \cite[][]{dumusque17}. These time series included the contribution of active regions as well as oscillations, granulation, and supergranulation. 
Among the results,  the analysis based on Gaussian processes performed the best in terms of planetary recovery. A criterion that allowed us to separate the good from the poor recovery was also defined; it is detailed below. 
A smaller blind test performed on a more limited sample of six time series by \cite{nelson20} focused on comparing Bayesian approaches and their robustness, but it assumed white Gaussian
noise (WGN) only to model the stellar contribution.  \cite{luhn22} implemented a systematic analysis based on a more realistic stellar model to compare the performance of different telescope and survey architectures in  characterising the mass of Earth analogues. The model included rotational modulation due to active regions, oscillations, and granulation based on Gaussian processes, but, importantly, did not include supergranulation, nor any source of long-term variability: However, these are  critical for the characterisation of these planets at long periods, and we found that they represent an important contribution to the RV variability, even after binning \cite[][]{meunier19e,meunier20b}. In addition, Luhn and collaborators mostly compared the amplitudes of the signal and computed the signal-to-noise ratio, but did not inject  simulated planet signals to test the detection capabilities directly. \cite{zhao22} compared many methods on a few RV time series from the EXtreme-PREcision Spectrograph (EXPRES), but also compared the root mean square (hereafter rms) of the RV residuals alone, without injecting any planet and therefore without studying the impact of the method on the planetary signal itself.

Current solar time series such as the one  obtained with
the High Accuracy Radial velocity Planet Searcher for the Northern hemisphere \cite[HARPS-N,][]{dumusque21} allowed us to  perform tests on planet-free time series (after the RVs of the Solar System planets, which are very well known, were removed) that were realistic enough.
The solar time series includes all processes, even  possible processes that have not been identified yet. However, it is only one such series, for the Sun seen edge-on, and for a limited duration.
 A complementary approach therefore is using synthetic time series because they allow us to consider a very large data set and a temporal coverage that is as long as needed. In addition, they allow us to consider various spectral types with different activity levels and inclinations. In both cases, it is possible to control the planetary signal that is injected, allowing for blind tests.
Here, we  perform  blind tests focusing on the impact of magnetic regions due to spot and plage contrasts and to the inhibition of the convective blueshift in plages, also based on a very large set of synthetic time series, assuming complex activity patterns.
 This type of simulations based on a complex activity pattern is much more realistic than simulations based on simple patterns (e.g. one or two spots), such as in \cite{desort07}, \cite{boisse12}, or \cite{dumusque14}, which are made to understand the behaviour and impact of individual physical effects
 \cite[e.g.][]{dravins21a,dravins21b}.

Our objective in this paper is therefore to perform massive blind  tests of planet detection and mass estimation on a very large set of synthetic time series, in which active regions (ARs) represent the dominant contribution, based on the very large set of time series produced in \cite{meunier19}, hereafter Paper I. A preliminary analysis of these time series in \cite{meunier19b} showed that the detection of Earth-like planets in the habitable zone of solar-type stars would be extremely difficult, even with many nights of observations over a long time basis. The comparison was based on a simple criterion derived from \cite{dumusque17}, however, and not on a blind test with injected planets.
In addition to this dominant contribution, we consider in the present work the contribution of oscillations, granulation, and supergranulation as in \cite{meunier19e,meunier20b}, to describe a more complete picture of the processes. We then explore two categories of blind tests. The first category aims to  quantify the precision of the estimation of the mass of planets that are detected via transit photometry, using RV data. We wish to compare the mass estimation performance 
with the mass uncertainty targeted by the RV follow-ups of PLAnetary Transits and Oscillations of stars (PLATO)  transit detections (10\%). 
The second category of tests focuses on blind searches of exoplanets in RVs as would be performed for example in a large survey. In both cases, we focus on Earth-like planets in the habitable zone around their host stars. We consider main-sequence stars of moderate activity level as in Paper I (i.e. not applicable to very young stars, but still covering a wide range of activity levels) between spectral types F6 and K4. 
In addition, we consider a correction  of the contribution of stellar activity, consisting of subtracting a model of stellar activity from the synthetic RV time series. The time series produced in Paper I includes $\log R'_{HK}$ time series, which is the main activity indicator  considered here. We  use a non-linear relation between RV and $\log R'_{HK}$ as well as a dependence on cycle phase due to the properties between these two variables discovered in \cite{meunier19c}. As a side result, we compare different correction methods.

The outline of the paper is the following.
The methods are presented in Sect.~2. We describe how the synthetic time series of stellar activity are built and how the injected planet was modelled. The standard correction for stellar activity applied throughout the paper is outlined. Sect.~3 presents the mass estimation performance of the RV follow-up of transit detections as a function of spectral type and planetary mass, and we also compare it with several other methods. Sect.~4 describes the method and the result of a full blind test allowing us to derive detection rates and false-positive rates. Finally, the residuals are analysed in Sect.~5 to identify where most of the improvement should be made, and we conclude in Sect.~6.


\section{Methods}

In this section, we first describe the model we used to describe the stellar contribution, which is mostly due to spots and plages, and then we present the planetary model. A brief overview of the approach is then provided.

\subsection{Modelling stellar activity and instrumental and photon noise}

To take the impact of magnetic activity into account, we used the large number of  RV  and chromospheric synthetic time series  that we described in detail in Paper I. 
These simulations are  based on a complex solar-like distribution of spots and plages on the stellar surface. Magnetic structures are generated based on various laws describing the lifetime and size distributions (solar parameters), butterfly diagram (distribution in latitude over the cycle, with different maximum latitudes in addition to the solar one), and various realistic rotation periods, activity levels, and cycle amplitudes, adapted from solar parameters \cite[][]{borgniet15} that were extrapolated to other stars based on observations whenever possible, for example based on the rotation-activity level from \cite{mamajek08}. The RV variability is then due to two physical processes: 1/ the contrast of spots and plages, and 2/ the inhibition of the convective blueshift in plages, as summarised in Table~\ref{tab_proc}. These contributions are hereafter named AR (active regions) because they are both due to magnetic structures, spots, and plages. 
We first describe the production of time series representing complex activity patterns. To be more realistic, we also included the contribution due to oscillations, granulation, and supergranulation, as well as a white gaussian noise (WGN) representing the uncertainty on the RV measurements due  to photon noise and instrumental effects, described in the following sections. The processes we took into account are summarised in Table~\ref{tab_proc}.

\subsubsection{Magnetic activity}

\begin{table*}
\caption{Processes taken into account in the RV synthetic time series}
\label{tab_proc}
\begin{center}
\renewcommand{\footnoterule}{}  
\begin{tabular}{lll}
\hline
Abbreviation & Processes & Method \\ \hline
AR & Spot and plage intensity contrast & Generation of structures with complex \\
   &   &   activity patterns (Paper I) \\
    & Inhibition of convective blueshift in plages & Same plage structures \\
OGS & Oscillations & Law from\cite{harvey84}\\
     & Granulation & Idem \\
      & Supergranulation & Idem\\
WGN & Poisson + read-out & Based on measured uncertainties (ESO archive)\\
\hline
\end{tabular}
\end{center}
\tablefoot{The AR contribution is described in more detail in Paper I (see Sect.~2.1.1) and extrapolated from the work of \cite{borgniet15} and from the OGS contribution in \cite{meunier20b}; see Sect.~2.2.2. The WGN contribution mostly includes the photon noise (Poisson) based on ESO (European Southern Observatory) archival data.
}
\end{table*}

\begin{figure}
\includegraphics{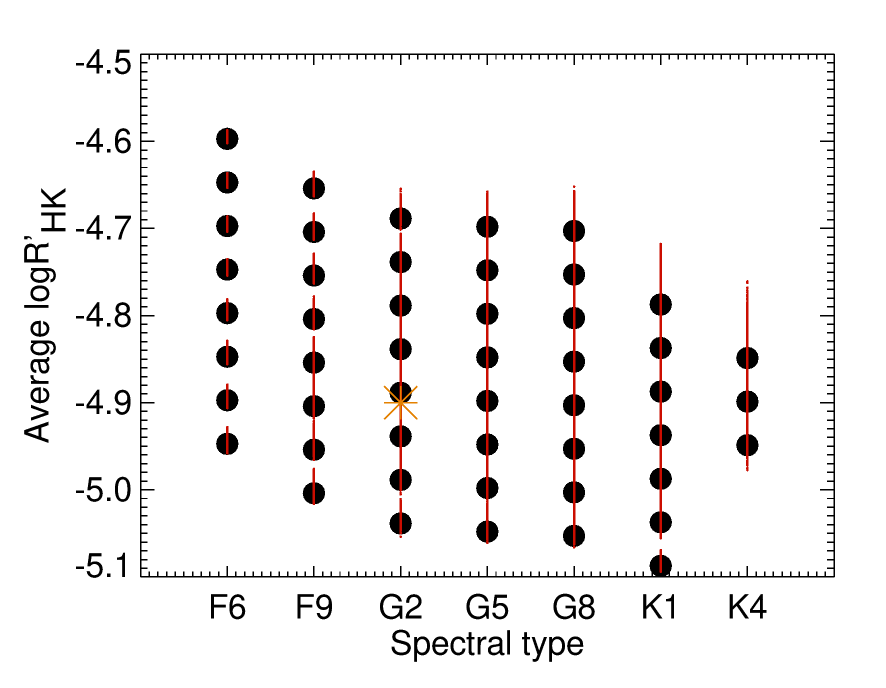}
\caption{
$\log R'_{HK}$ vs. spectral type in our grid of simulations of the magnetic activity contribution (Paper I). Large filled circles correspond to the averaged targeted $\log R'_{HK}$ in the grid, and the actual average (over each time series) $\log R'_{HK}$ in all simulations (the dispersion of the dots is due to each realisation and inclination) is represented in red. The lower envelope corresponds to the observed $\log R'_{HK}$ as a function of spectral type \cite[e.g.][]{mittag13,borosaikia18}, and the upper envelope was chosen in Paper I to correspond to the limit between stars that are mostly dominated by plages in photometry (like the Sun) and stars that are spot dominated following \cite{radick18}. The average position of the Sun is shown as an orange star. 
}
\label{logrphk}
\end{figure}

\begin{table}
\caption{Typical RV rms (in m/s) in the synthetic time series }
\label{tab_rms}
\begin{center}
\renewcommand{\footnoterule}{}  
\begin{tabular}{llll}
\hline
Process & F6 & G2 & K1  \\ \hline
AR & 0.13-9.20    &  0.1-6.8 & 0.04-5.03  \\
OGS &  0.37-0.39  &  0.31-0.37  (*)  & 0.26-0.30\\
WGN & 0.09, 0.17, 0.45    &  0.09, 0.17, 0.45 & 0.09, 0.17, 0.45\\
\hline
\end{tabular}
\end{center}
\tablefoot{Examples of typical RV rms in our time series that is due to the different contributions: AR, which is the focus of this paper \cite[][]{meunier19,meunier19b}; OGS, typically for a low level of granulation and supergranulation \cite[][]{meunier20b}; and WGN, described in Sect.~2.1.3  (shown by the asterisk). In addition, tests were made for G2 stars alone with a medium level of supergranulation, corresponding to typically 0.72 m/s for the OGS contribution. 
}
\end{table}

\begin{table*}
\caption{Configurations}
\label{tab_config}
\begin{center}
\renewcommand{\footnoterule}{}  
\begin{tabular}{lllllll}
\hline
Number & Configuration & Spectral & Planet & Noise & Yearly & SG \\
     &                & type & mass & levels & gap & \\
     \hline
1 & MAG+OGS+WGN  & {\it All} & {\it 1-4~M$_{\rm Earth}$} & {\it All} & {\it 4 months} & {\it low } \\
   &   &   G2 &  1-2~M$_{\rm Earth}$ & 0.09~m/s&  6 months & low\\   
      &   &   G2 &  1-2~M$_{\rm Earth}$ & 0.09~m/s&  4 months & high\\   
2 & AR+WGN & G2 & 1-2~M$_{\rm Earth}$ & 0.09~m/s & 4 months & low  \\
3 & OGS+WGN & G2 & 1-2~M$_{\rm Earth}$ & 0.09~m/s& 4 months  & low \\
4 & WGN &  G2 & 1-2~M$_{\rm Earth}$ & All  &  4 months  & low \\
5 & AR+OGS &   G2 & 1-2~M$_{\rm Earth}$ & 0.09~m/s & 4 months &low  \\
6 & AR &  G2 & 1-2~M$_{\rm Earth}$ & 0.09~m/s & 4 months  &low \\
7 & OGS &  G2 & 1-2~M$_{\rm Earth}$ & 0.09~m/s & 4 months  &low\\
\hline
\end{tabular}
\end{center}
\tablefoot{ Configurations tested in the paper. The main configuration, including all effects (AR, OGS, and WGN), indicated in italics, is  number 1 (reference configuration in the text). The other configurations (additional configurations for numbers 1 and 2 to 7) are tested in Sect.~3.5 and 3.6 alone. }
\end{table*}

The parameters cover a wide range of stellar activity levels that correspond to relatively old (typically older than 1 Gyr) main-sequence stars for different spectral types (F6-K4), as illustrated in Fig.~\ref{logrphk}. An exhaustive analysis of the RV jitter due to these contributions was made in \cite{meunier19b}.
We consider seven different spectral types between F6 and K4 below. Time series were produced with two levels of spot contrast, the first level (denoted by ($\Delta T_{\rm spot1}$)  corresponding to a solar spot contrast,  and the other level, denoted $\Delta T_{\rm spot2}$, to the upper limit of the spot contrast from the sample of stars reported in \cite{berd05}. 
Since the spot contrast directly impacts  the amplitude  of the short-term  variability, we considered both in the present analysis. 
All time series were generated for ten  inclinations between 0$^\circ$ and 90$^\circ$, with a step of 10$^\circ$. 
The typical RV rms values for G2 and K1 are shown in Table~\ref{tab_rms}.  The convective blueshift decreases towards K stars, but they are also more active stars on average. The RV jitter therefore covers a wide range of values for each spectral type (as in indicated in Table~\ref{tab_rms}), with a trend towards a lower signal for K stars. The wide range in activity variability for each spectral type strongly impacts the distributions, with no strong trend due to this dispersion.  

\subsubsection{Oscillations, granulation, and supergranulation}

As oscillations, granulation, and supergranulation (hereafter OGS) significantly impact the characterisation and detectability of low-mass planets at long periods \cite[][]{meunier19e,meunier20b}, we built time series that were added to the AR time series described above, considering a relatively optimistic long equivalent exposure time of one hour because it allows a good reduction of the granulation signal (typically by a factor of two), while averaging over longer timescales is not significantly more efficient \cite [][]{meunier15,sulis22b}. This equivalent exposure time of one hour can be the sum of several exposures because it is usually not possible to perform  exposures of this duration directly. We considered them to be consecutive without gaps, neglecting the readout time and considering the ideal case of a readout noise of one exposure. The impact on the rms is a few percent of the granulation signal for the brightest stars at most (which would require short exposure times) and below 0.3\% for the supergranulation (hereafter SG) signal. 

The RV time series were generated based on the power spectrum laws from \cite{harvey84}, as in \cite{meunier20b}: Granulation amplitudes were calibrated based on observations \cite[][]{palle99} and 3D HD simulations \cite[][]{sulis20}, and  supergranulation amplitudes based on previous simulations \cite[][]{meunier15}.  Most computations were performed assuming a low level of granulation (0.4 m/s for a G2 star) and a low level of supergranulation (0.28 m/s). We show in the following sections that the activity signal (which is dominant) combined with this low OGS level leads to a poor detection performance. It is reasonable to use this level in most computations.

In addition, a few tests were also performed with a supergranulation level of $\sim$0.7 m/s (for a G2 star), which corresponds to the medium level obtained in \cite{meunier15} and to the highest level studied in detail in \cite{meunier20b}. It is compatible with the supergranulation level observed by \cite{palle99} and with the day-to-day RV dispersion in the solar observation with HARPS-N \cite[][]{dumusque21}, although its precise amplitude remains uncertain because these dispersions may only be upper limits. 
The typical RV rms values for G2 and K1 are shown in Table~\ref{tab_rms} and are compatible with recent granulation levels estimated with ESPRESSO (Echelle SPectrograph for Rocky Exoplanets and Stable Spectroscopic Observations) observations for an F7V and G0V stars \cite[][]{sulis22b}. The corresponding rms values for all spectral types can be found in \cite{meunier20b}:  The RV jitter decreases from F to K stars.

 \subsubsection{White Gaussian noise}
 
In addition to the AR and OGS complex signatures, we added a WGN contribution to account for photon noise. To be consistent with the previous section (an exposure time of one hour for the OGS contribution), we  chose the amplitude of this noise as follows. First, we considered HARPS-like instruments. An instrument such as ESPRESSO is more stable and leads to lower RV uncertainties. However, our objective is to focus on low-mass planets in the habitable zone of solar-type stars (i.e. long-term observations), which will probably be limited with ESPRESSO because previous works \cite[e.g.][]{meunier20b,luhn22} showed that  a large number of observations over a long time will be necessary. 
We therefore estimated that it was more realistic to consider that these observations will be performed with HARPS or an instrument of similar class, for example with HARPS3 \cite[][]{thompson16}. To consider a realistic level of noise, we used the RV uncertainties provided by the HARPS DRS (Data Reduction Software) for the large sample of FGK stars analysed in \cite{meunier22}. We analysed these uncertainties as a function of exposure time and magnitude and extrapolated them to an equivalent exposure time of one hour (hence neglecting possible additional readout noise caused by multiple intermediate exposures, which is reasonable assuming that the spectra are acquired with a very good signal-to-noise ratio), providing a range of values:
We find a WGN of 0.09~m/s for a V=4 magnitude, 0.17~m/s for a  V=7 magnitude, and 0.45~m/s for a V=10 magnitude. The 0.09~m/s could also correspond to weaker stars observed with ESPRESSO. We considered these three levels, which are the standard deviations of the WGN added to the data, although our reference level is a WGN of 0.09~m/s in the following. For a given star, the exact noise level might be different, for example depending on the reading overheads and on its flux at the time of observation, because the added noise corresponds to a certain flux and therefore to a specific photon noise. We did not include any dependence on spectral type because there was no clear trend from this analysis. We may expect for example K stars to provide better RV uncertainties because there are more spectral lines. Because we consider a wide range of values, this also includes this type of impact.
This analysis does not take any additional instrumental effects into account, which is not easy to model realistically in a general case and would probably degrade the performance further, so that our values are optimistic overall.

\subsubsection{Summary: Construction of the stellar time series}
 

For each spectral type, we selected the AR time series in the sample described in Paper I with a duration longer than ten years. For a given inclination, this selection therefore provides between 243 and 648 time series, depending on spectral type (times ten inclinations and two spot contrasts), for a total of 79020 series. For each time series (corresponding to a given inclination between 0$^\circ$ and 90$^\circ$), we selected a random temporal sampling as follows: We considered a ten-year duration; in most cases, we considered gaps of  four consecutive months each year (tests with a six-month gap were also performed); 1000 nights of observations were then randomly spread over the remaining available days as in \cite{meunier20}; and finally, we considered one hour of observation per night, taken consecutively.
For each time series, the OGS and WGN contributions were then added.  This was done for the two levels in spot contrasts described in Sect.~2.1.1. 
Several hundred time series for each configuration were also generated separately for the AR, OGS, and WGN contributions to compare the relative impact of the different contribution to RV variability, including a higher level of OGS signal. 

In addition, the plages generated in the AR simulations were  used to produce  synthetic $\log R'_{HK}$ time series based on laws relating plage sizes and chromospheric emission from \cite{harvey99}. A WGN with a level similar to that observed for HARPS data and FGK stars was added, as determined in \cite{meunier22}, corresponding to a level of 5 10$^{-4}$ on the $\log R'_{HK}$  values. 
These $\log R'_{HK}$ time series are used in the following (Sect.~2.3) to correct the RV time series for stellar activity.  The spots and plages are also used to produce  photometric time series \cite[analysed in][]{meunier19d}, which are used in one of the tests. 

The resulting time series therefore include a large panel of stellar contributions. The main contribution that is not included is due to meridional circulation \cite[][]{makarov10b,meunier20c}. \cite{meunier20c} found a variability that was globally correlated with the cycle ($\log R'_{HK}$) for stars seen edge-on (with an amplitude from a few 0.1 m/s to 1.7 m/s depending on mass and rotation rate) and anticorrelated for stars seen pole-on (with possible amplitudes up to 4 m/s), although with a poorly defined possible phase shift. Even though we expect this process to lead to significant RV variability, it was not included in this analysis because the relation with the cycle seems to be complex and remains to be better understood before it is considered in such a systematic approach.

\subsection{Modelling the planet}

To the simulated time series described in Sect.~2.1, we added a planetary RV signal.  We focused on planets  with masses between 1 and 4 M$_{\rm Earth}$, orbiting in the habitable zone. For simplicity, we assumed the orbits to be circular. The planet was in addition assumed to orbit in the equatorial plane of the star as in \cite{meunier19e} and \cite{meunier20b}; therefore, the inclination here is the same for the star and the orbital plane of the planet.
We  compared the performance based on this assumption with an inclination distribution between the orbital plane and the equatorial plane in an astrometric study \cite[][]{meunier20}, but found little impact.
We considered systems with only one planet. The habitable zone was defined as in our previous analysis \cite[][]{meunier19b} following the prescription of \cite{kasting93} based on luminosity \cite[][]{jones06,zaninetti08}.  We therefore followed the classic definition, corresponding to the range of distances in which liquid water could be present, and only luminosity effects were taken into account. The inner side corresponds to a runaway greenhouse effect, which would imply the evaporation of all the surface water, and the
outer side is the maximum distance corresponding to a temperature of 273 K in a cloud-free CO2 atmosphere. The shortest period ranges from 179 to 410 days (from K4 to F6 stars) and the longest period ranges from 502 to 1174 days.

Finally, we considered two categories of analysis. First, RV observations can be performed as a follow-up of a transit candidate. In this case, we only considered edge-on simulations, given the assumption considered on the inclination between the planet orbital plane and the star equatorial plane, and the period is precisely known. Second, we considered blind searches. In this case, all stellar inclinations and therefore all orbit inclinations with respect to the line of sight have to be considered.

\subsection{Setting of this study}
\label{sec23}

\subsubsection{Correction method}

Without any correction of the signal due to activity, the  signals of  1-4 M$_{\rm Earth}$ planets cannot be detected. It is therefore necessary to apply a noise model. In most results presented here, we considered a correction of the stellar signal based on the $\log R'_{HK}$ time series. 
This activity indicator primarily allows us to correct for the contribution due to the inhibition of the convective blueshift in plages because both  the RV signal due to this process and $\log R'_{HK}$
are strongly correlated with the plage filling factor, which is not the case of the other contributions. 
Correcting for the inhibition of the convective blueshift is critical because it is the main contributor to the long-term variability in our simulations, and we  focus on planets in the habitable zone and therefore at long periods as well. 
However, we did not consider a simple linear correlation between $RV(t$) and  $\log R'_{HK}(t)$, which presents limitations \cite[e.g.][]{meunier13}, but instead, we considered a slightly more complex model to take the hysteresis discovered between the two variables into account \cite[][]{meunier19c}. This non-linearity is due to the combination of two facts. First, the activity pattern is not always in the same latitude range over the cycle, that is, on long timescales (butterfly diagram). Therefore, the average position of the plages on the disk varies with time and corresponds to different average centre-to-limb distances. Second, both processes (inhibition of the convective blueshift and chromospheric emission) suffer from different projection effects. There is therefore a departure from the linear correlation that should be taken into account, with a non-linear dependence of the RV on $\log R'_{HK}$ as well as a dependence on cycle phase. One objective of this paper therefore is to test this model and to compare it with other approaches (see below). 
We modelled the RV due to activity as

\begin{equation}
\label{eqmodel}
\begin{split}
 RV_{AR}(t) = A \cdot (1+B \cdot \log R'_{HK}(t) + C \cdot (\log R'_{HK}(t))^2) \\
\times ( 1 + D \cdot \phi (t) + E \cdot \phi (t)^2) + F
 \end{split}
,\end{equation}

where $\phi$ is the cycle phase. Parameters 
 $A$, $B,$ and $C$ characterise a second-degree polynomial relation between $\log R'_{HK}$ and RV. Parameters $D$ and $E$ characterise a departure from the relation based on $\log R'_{HK}$, using a second-degree polynomial relation between the phase $\phi$ and RV. Parameter $F$ is a constant.
  In the scope of this paper, we investigated the simple (quadratic) non-linear model above, but more complex models can be studied in the future, for example based on several activity indicators \cite[e.g.][]{perger23}. 
 The phase $\phi$ was estimated after estimating the cycle period for each realisation from the $\log R'_{HK}$ time series. The method is described in Appendix \ref{secB}. 
 The true period is clearly known from the simulation parameters, but it was not used here to place ourselves in realistic conditions. The fitted cycle period is very good for the lowest periods, and it is noisier for the longest periods because they are not well sampled over a ten-year coverage (the range of periods is given in Sect.~2.2).
 
\begin{figure}
\includegraphics{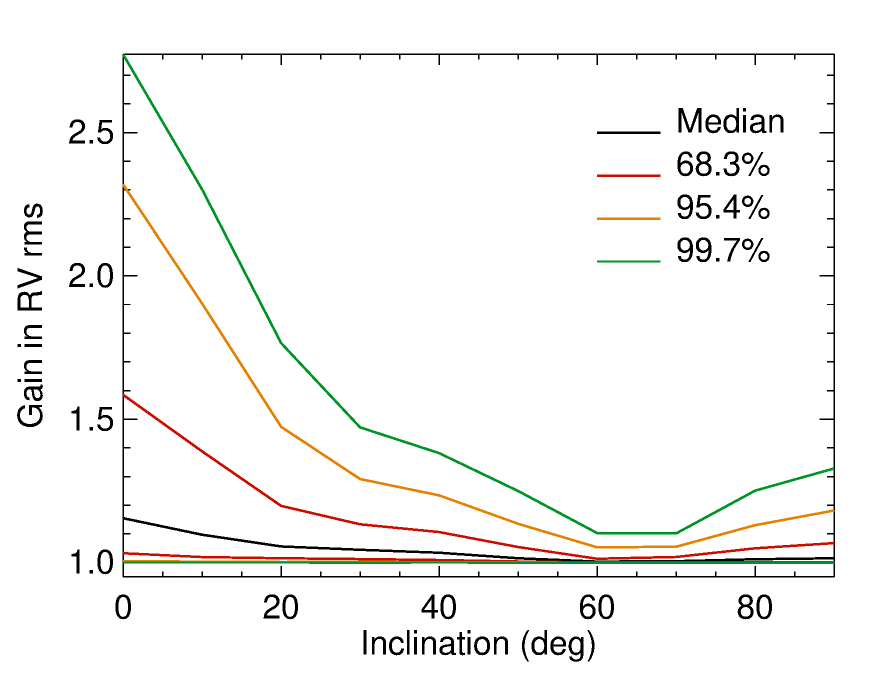}
\caption{
Gain in RV rms on the residuals brought by the model of  equation~\ref{eqmodel} compared to a simple linear model in $\log R'_{HK}$ vs. stellar inclination. The gain is defined as the rms of the residuals after a correction for the linear model divided by the rms of the residual after correction for the non-linear model. The black line indicates the median  values over all realisations for the seven spectral types, and the colours correspond to different percentiles: 68.3\% (red), 95.4\% (orange), and 99.7\% (green).
}
\label{gainrms}
\end{figure}

\begin{figure}
\includegraphics{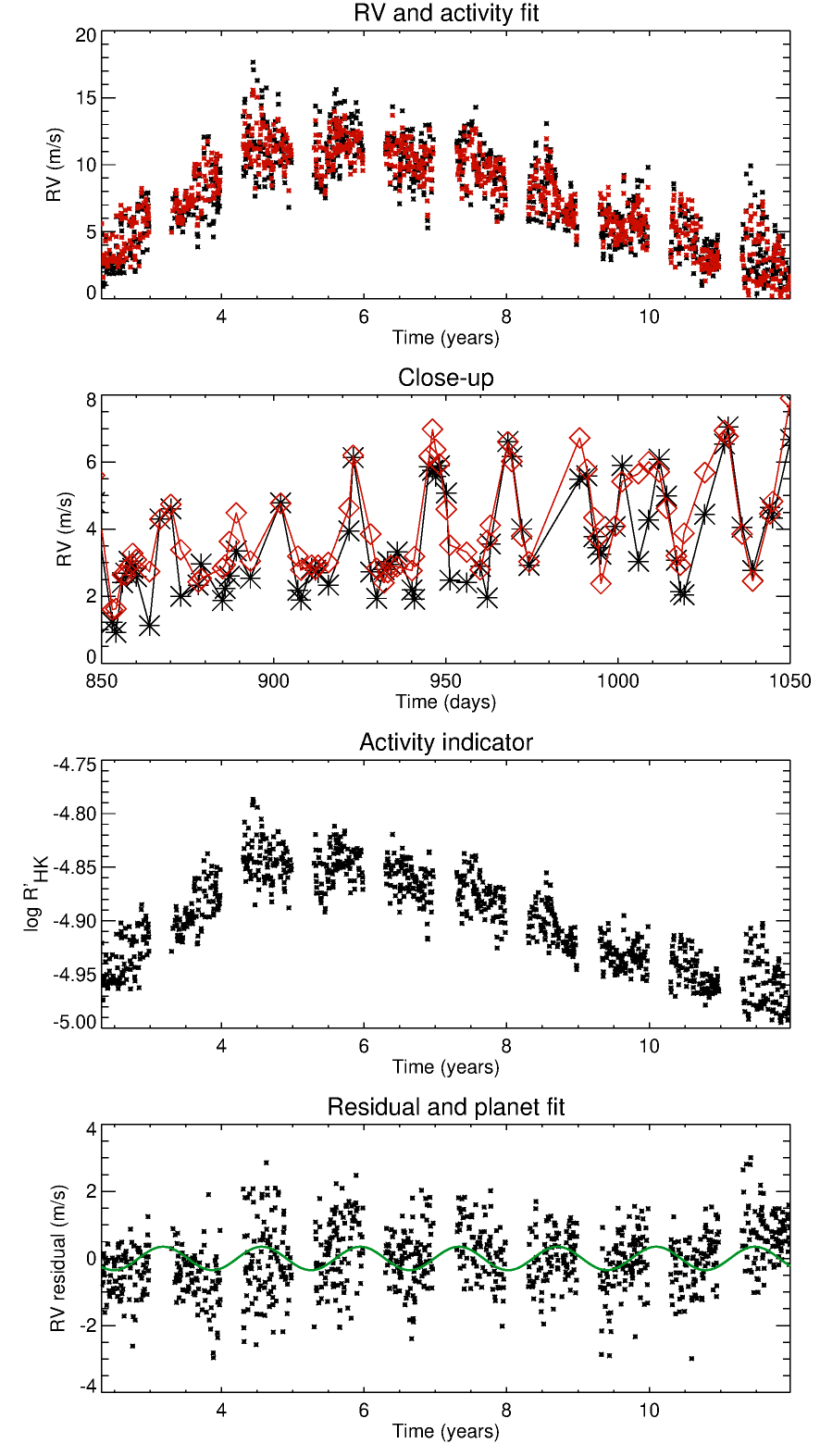}
\caption{
Example of RV time series (upper panel) and close-up in time (second panel). The simulation is shown in black, and the AR model according to equation~\ref{eqmodel} is shown in red. 
The  third panel shows the $\log R'_{HK}$ time series, 
and the last panel shows the residuals after correction. The true planetary signal (4~M$_{\rm Earth}$, middle of the habitable zone) is shown in green for comparison.   
}
\label{excorr}
\end{figure}

\begin{figure}
\includegraphics{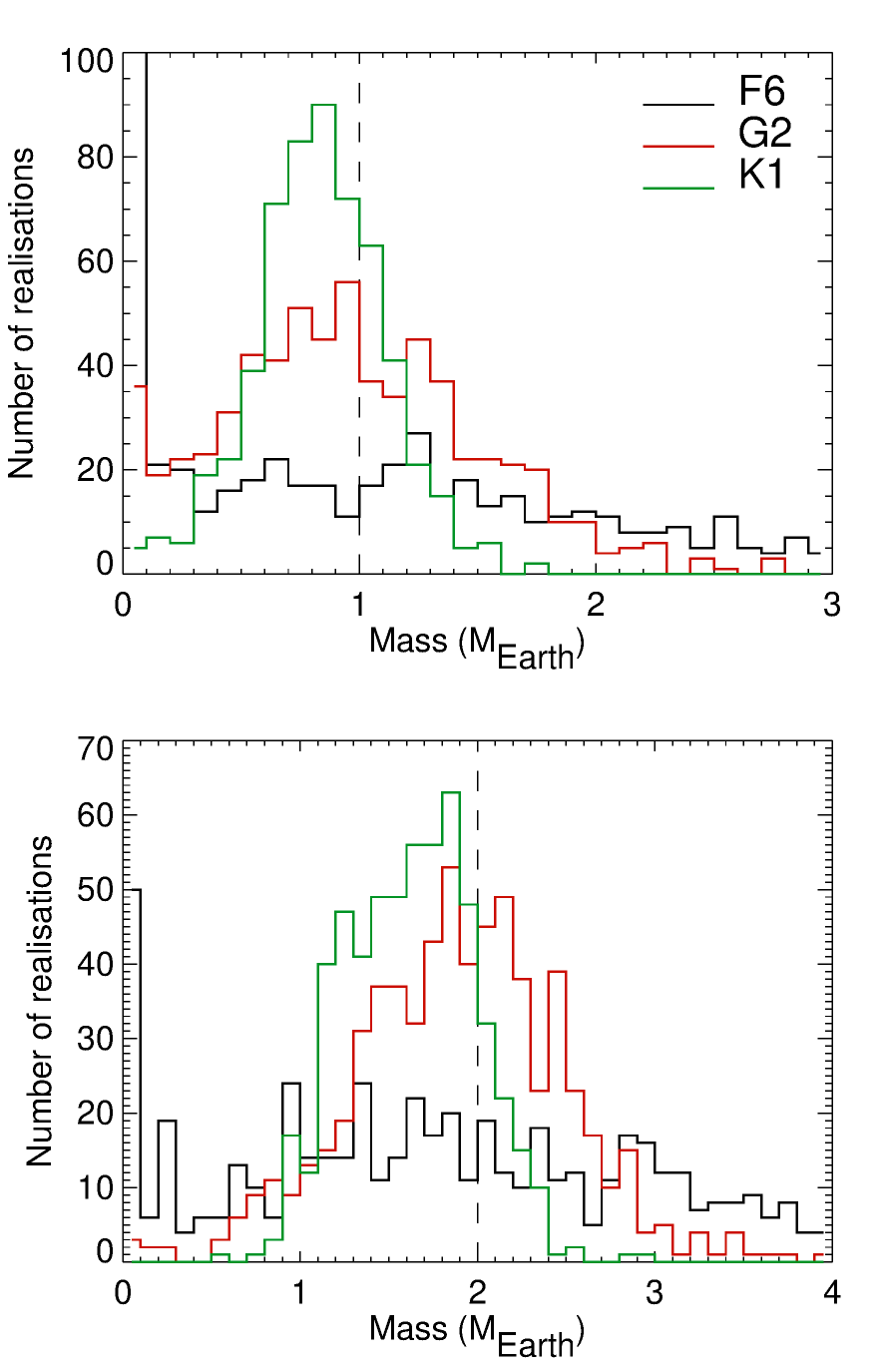}
\caption{
Selection of fitted mass distributions in the follow-up blind tests for 1 M$_{\rm Earth}$ (upper panel) and 2 M$_{\rm Earth}$ (lower panel) for PHZ$_{\rm mid}$, $\Delta T_{\rm spot1}$, a WGN noise level of 0.09 m/s, and for different spectral types: F6 (black), G2 (red), and K1 (green). The true mass is indicated by the vertical dashed line. 
}
\label{exdist}
\end{figure}

\begin{figure}
\includegraphics{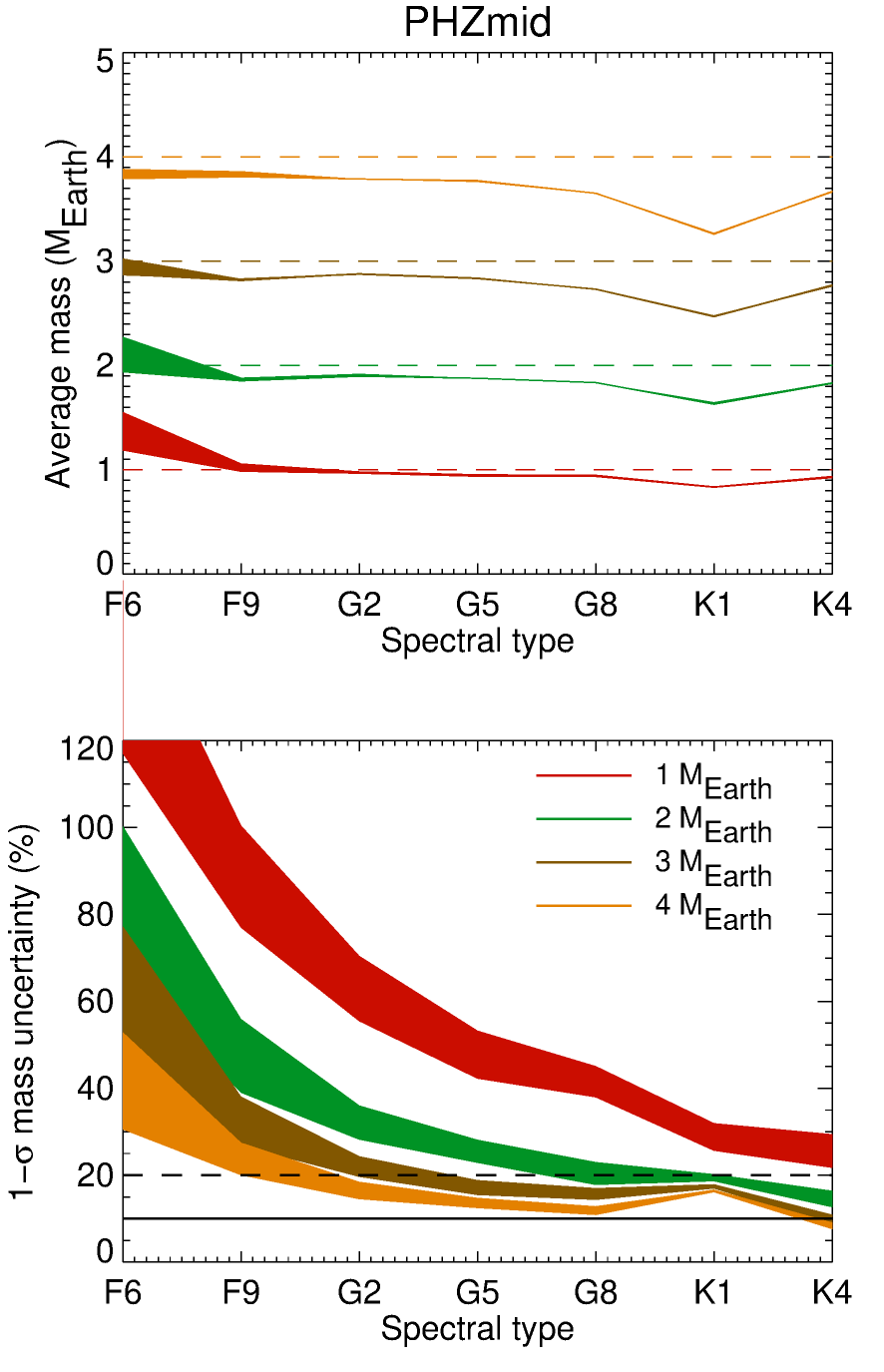}
\caption{
Averaged mass (upper panel) and 1$\sigma$ mass uncertainty (lower panel) vs. spectral type for PHZ$_{\rm mid}$ and for four  masses: 1 M$_{\rm Earth}$ (red), 2 M$_{\rm Earth}$ (green), 3 M$_{\rm Earth}$ (brown), and 4 M$_{\rm Earth}$ (orange). The thickness corresponds to the range covered by the spot contrast. The solid horizontal line indicates the 10\% objective for PLATO, and the dashed line shows an indicative level of 20\%. 
}
\label{vsts}
\end{figure}

 The results of the performance on the cycle period are detailed in Appendix \ref{secB}. The RV residuals were computed by subtracting the AR stellar contribution  estimated from this model and were then used for analysis. 
Fig.~\ref{gainrms} shows the gain in rms on the residuals compared to a linear dependence on $\log R'_{HK}$: The gain is defined as the rms of the residuals after a correction for the linear model divided by the rms of the residual after correction for the non-linear model. For each inclination, we considered the list of gain values and represent the different percentiles The gain is always larger than 1. It is maximum  for stars seen pole-on and minimum around 60-70$^\circ$, as expected from \cite{meunier19c}, because the above-mentioned hysteresis also follows this dependence on inclination. The gain for a star seen edge-on is 30\% at most. 
 Figure~\ref{excorr} shows an example of the correction, with a 4~M$_{\rm Earth}$ planet injected in the middle of the habitable zone, as well as the typical residuals that can be obtained. In this example, the rms of the residuals after correction for the activity signal is 0.93 m/s, and 0.89 m/s after the removal of the planetary signal. We recall that this approach aims to subtract most of the contribution due to the convective blueshift in plages, but it cannot remove the other stellar contributions (spot and plage contrasts, for which we expect some residuals close to the rotation period), nor the OGS and WGN noise. 

\subsubsection{Blind tests}

For the RV follow-up of a transit candidate, we considered that the presence of the planet is known, as are its period and phase. We fit the planet mass in order to evaluate the uncertainty on the mass depending on spectral type, stellar activity level, and planet mass. 

In the case of a blind search, a planet is injected in some simulations and not in others. In this case, the periodogram was computed, and the highest peak 
was compared to a false-alarm probability (fap) level to decide whether this was a detection (peak above the fap level) or not (peak below the fap level). If there was a detection, a fit was performed on the RV residuals to retrieve the planet parameters. By comparison with the true parameters, 
detection rates and false- positives rates can be estimated. More details about the protocols are provided in Section 3 and 4. 

In addition, a similar analysis was performed on a few time series corresponding to the AR, OGS, and WGN contributions separately to compare the results with the complete time series.  The residuals can also be studied to understand the limitation of the correction procedure and to identify the type of improvement that should be performed: this is done in Sect.~5. Finally, we also tested other correction techniques in the case of the follow-up protocol, including denoising methods based on various assumptions, a variant of the FF' method of \cite{aigrain12}, and Gaussian processes. We find that none of these approaches significantly improve the performance and some are doing worse. They are therefore presented and discussed in Appendix A.

\section{Radial velocity follow-up of a transit detection}
\label{sec3}

This section is devoted to the mass estimation based on RV observations following a photometric transit detection, mimicking the follow-up strategy envisioned for the PLATO Earth-like planet candidates. We first describe the protocol adopted for these blind tests. Then, we study the dependence of the mass uncertainty on spectral types and activity levels. The impact of the different contributions as well as the impact of the yearly gap duration are then studied.

\subsection{Method}

For each RV simulation, we added the signal of a planet in the habitable zone as described in Sect.~2.2, with three possible orbital periods: the inner side of the habitable zone (PHZ$_{\rm in}$), the middle (PHZ$_{\rm mid}$), and the outer side (PHZ$_{\rm out}$). Each series was then corrected for the stellar contribution as described in Sect.~2.3 (model in equation~\ref{eqmodel}), and a planetary signal without eccentricity  was fitted in the residuals with known period and phase (from transits).

We then converted the RV amplitude into a mass (based on the Kepler laws) that was then compared to the true mass (1, 2, 3, and 4 M$_{\rm Earth}$). The properties of the residuals (after removal of the planetary fit in most cases) are studied in Sect.~5. Most of the results correspond to the reference configuration (indicated in italics in Table~\ref{tab_config}), which includes all  effects (AR, OGS, and WGN).  
In the following, the uncertainty on the estimated mass for a given set of realisations is derived from the distribution of fitted masses. For example, all realisations corresponding to a given spectral type, a planetary mass, a spot contrast, and a position in the habitable zone lead to a 1$\sigma$ level estimated  from this distribution. 
We used this approach because the least-squares fits were performed with a gradient descent and provided under-estimated uncertainties, while an MCMC (Markov chain Monte Carlo) approach is impractical here given the large number and the length of the  time series ($\sim$ 50000). The uncertainties  therefore correspond to an average uncertainty for a given set of realisations of the parameters, typically, for a given spectral type and planetary mass. This Monte Carlo approach provides a direct estimate of the uncertainty for a sample of stars (e.g.  stars of the same spectral type and a given planet mass), allowing us to compare this with the PLATO objective. This does not prevent the possibility that certain stars have a lower uncertainty, mostly due to different activity levels: this is taken into account in Sect. 3.4. \\

\subsection{Fitted masses}

Figure~\ref{exdist} shows the distribution of the fitted mass for a few examples, 1~M$_{\rm Earth}$ (upper panel) and 2~M$_{\rm Earth}$ (lower panel), the middle of the habitable zone, the low spot contrast, and a selection of three spectral types. For F6 stars, the mass is very poorly determined, with a very wide distribution of  fitted masses. The mass distribution for G2 stars is closer to a Gaussian, except for a truncation at M=0 and an excess of realisations in that mass bin. Finally, for K1 stars, the distribution is Gaussian-like with a lower dispersion of about 20\%.  
The half-width at half maxima of a Gaussian distribution fitted on the G2 and K1 distributions for the 1~M$_{\rm Earth}$ (resp. 2~M$_{\rm Earth}$) are 0.71 and 0.30~M$_{\rm Earth}$ (0.65 and 0.46~M$_{\rm Earth}$). 
We note an offset between the true mass and the peak of the distribution, the mass being slightly underestimated, although this offset is within the uncertainties. 

The same analysis was performed for the high spot contrast. 
The distributions are similar, but the dispersion is higher because the spot contribution, which is mostly at the rotational timescale, is not corrected for when using the use of the correlation with the $\log R'_{HK}$ (because the shape of the signal is different). In this case, the residuals are therefore higher than for the low spot contrast, leading to a poorer performance.

\subsection{Dependence on the star spectral type}

We now consider the results for all spectral types and planet  positions  inside the habitable zone for the reference configuration (Table~\ref{tab_config}) and for the WGN of 0.09 m/s. Figure~\ref{vsts} shows the average fitted mass  and 1$\sigma$ uncertainty on the mass versus spectral type for the four masses and PHZ$_{\rm mid}$. The results are very similar for the two other positions in the habitable zone, PHZ$_{\rm in}$ and PHZ$_{\rm out}$, with slightly lower and higher uncertainties, respectively.
The mass is usually slightly under-estimated, as noted above. 
The bias is within the uncertainties, however. They are shown in the upper  panel of Fig.~\ref{vsts}. The uncertainty is always higher than the objective of 10\% for the PLATO mission. 
For 1 and 2 M$_{\rm Earth}$, the uncertainty is always higher than 20\% \footnote{\cite{batalha19} also recommend this level for detailed atmosphere characterisations.}, except for K4 stars. This means that it will be very difficult to characterise planets like this around G stars. For 3 and 4  M$_{\rm Earth}$, the 20\% level cannot be reached for F stars, but it can be reached in the middle of the habitable zone for G and K stars (G2 and earlier for 4 M$_{\rm Earth}$). 
The difference in results with respect to spectral type could be due to the difference in stellar variability (the distribution of the RV jitter tends to be lower for K stars, but with a strong overlap; see Sect. 2.1.1), but also to different planetary contribution: We considered planets in the habitable zone, which varies with spectral type, so that the planetary amplitude in RV is  higher for K stars. To test the respective impact of the period and the activity level, we performed a similar computation using the G2 period (middle of the habitable zone) and stellar mass for the K4 activity time series (1~M$_{\rm Earth}$), and vice versa. We find that a significant fraction of the difference is due to the different period and stellar mass between spectral types, but that this is not sufficient to explain the results, so that the difference in stellar variability contribution also plays a role.

We obtain similar results when considering the other WGN  levels (0.17 and 0.45 m/s), and the differences are within the uncertainties. 
Hence, the precision on the mass does not strongly depend on the level of moderate WGN. 
Different levels are therefore not critical for the mass uncertainties. We recall that these values were chosen to correspond to one-hour exposure times, which also have the strong advantage of decreasing the granulation signal by a factor of two \cite[][]{meunier15}. 
The lack of sensitivity to the WGN level is probably due to the fact the signal is dominated by stellar activity. This does not mean that it is not an important factor, however, for two reasons. First, there is an impact when it is considered alone (i.e. in an ideal case; see Sect.~3.5), so that in a condition with a very low stellar contribution (e.g. if a high correction can be achieved) it will be critical. Second, methods based on more sophisticated approaches (e.g. line-by-line fitting) will require very good signal-to-noise ratios.

\subsection{Dependence on the star activity level}

 We considered different criteria to quantify  the impact of the activity level: the cycle amplitude, the average $\log R'_{HK}$, and the RV rms before and after correction. We expect the mass uncertainty to be lower for lower-activity stars. For the first three criteria, we see no strong trends, however. We attribute this result to the fact that  the RVs of stars with stronger variability are also more strongly corrected. 

However, interestingly, the rms of the residuals is strongly correlated with the mass uncertainty. The details are shown in Appendix~\ref{appAA}. An extrapolation of the trend allows a rough estimation of the  RV rms in the residuals that is necessary to reach an uncertainty of 10\%. 

In addition, we attempted to quantify this more precisely by using criterion C proposed in \cite{dumusque17}, which is defined as

\begin{equation}
\label{eqc}
{\rm C}=\frac{{\rm K}_{\rm pla} \sqrt{N_{\rm obs}}}{{\rm rms_{\rm res}}} 
,\end{equation}

where K$_{\rm pla}$ is the amplitude of the planetary signal, and rms$_{\rm res}$ is the rms of the RV time series after correction. C is a dimensionless number related to the signal-to-noise ratio of a single sinusoid (in the ideal case, where the planet frequency is exactly on the Fourier grid and the noise is white and Gaussian), but weighted by the number of observations (assuming a regular sampling with no gap): a high value of C should allow a detection, while a low value should not. 
All curves are above the 10\% level. A 20\% mass uncertainty corresponds to C typically in the 8-12 range, with targeted RV rms of the residuals ranging from around 0.2-0.4 m/s (1~M$_{\rm Earth}$) to more than 0.8 m/s (1~M$_{\rm Earth}$). This analysis provides in principle a practical order of magnitude of the typical rms that should be reached by other methods or models. However, there are limitations: This criterion
does not take any frequency dependence of the RV time series (coming from both the star and the planet) into account, nor the specific temporal coverage of the sampling.  In addition, the criterion does not guarantee that the alternative approach to modelling the stellar activity does not degrade the planetary signal, and blind tests such as the one performed in this paper are always necessary to verify that they do not.

\subsection{Impact of the different contributions}

\begin{figure}
\includegraphics{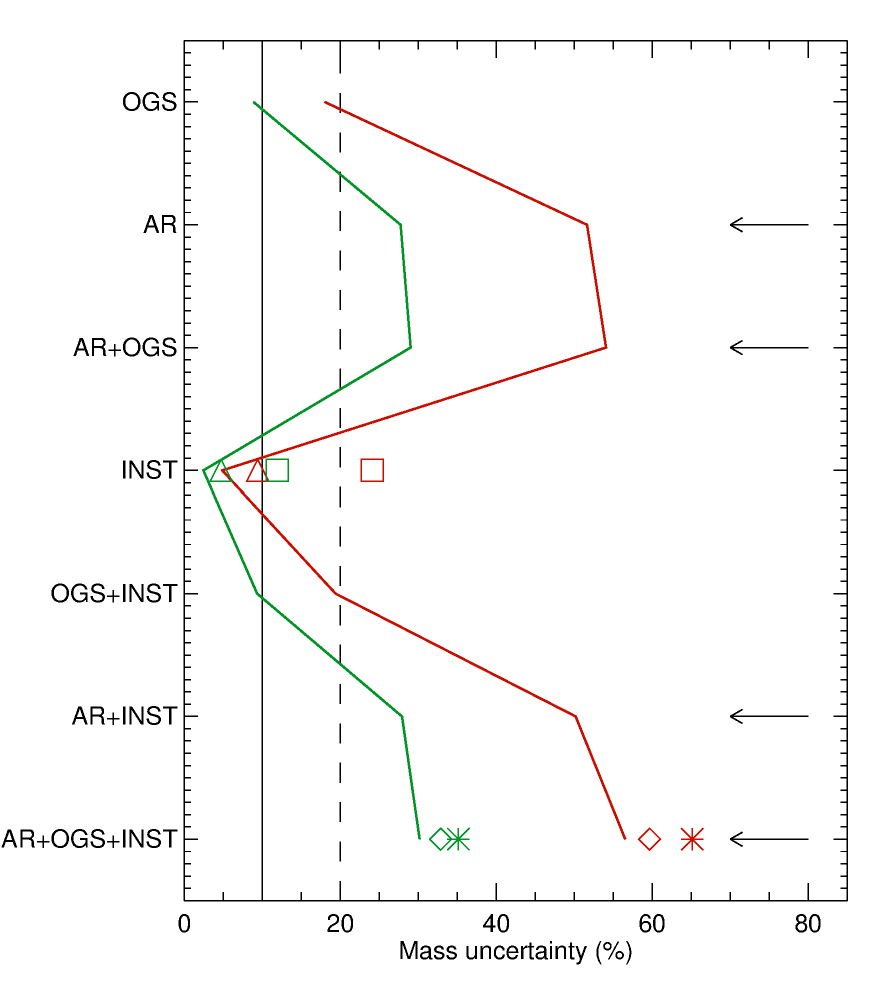}
\caption{
Mass uncertainty in the follow-up blind tests vs. configuration (see Table~\ref{tab_config} for details) for $\Delta T_{\rm spot1}$, 1 M$_{\rm Earth}$ (red), and 2 M$_{\rm Earth}$ (green) for the WGN of 0.09 m/s, G2 stars, and for PHZ$_{\rm mid}$. Arrows highlight configurations that include the AR contribution. Other symbols correspond to other specific configurations: a higher level of supergranulation (stars), a six-month gap instead of a four-month gap (diamonds), a noise level of 0.17 m/s (triangles),  and a noise level of 0.45 m/s (squares), with the same colour code for the mass (three identical symbols are used for all PHZ values to simplify the representation).
}
\label{contrib}
\end{figure}

For G2 stars, we  performed similar blind tests for configurations 2-7 summarised in Table~\ref{tab_config}, that is, when considering only one or two contributions in the AR, OGS, and WGN list. This was done for 1 and 2  M$_{\rm Earth}$ and the three periods in the habitable zone. No activity correction was applied for configurations 3, 4, and 7 because they do not include the AR contribution.  Figure~\ref{contrib} shows the mass uncertainty for $\Delta T_{\rm spot1}$ (the graph is very similar for $\Delta T_{\rm spot2}$). Configuration 1 is the reference configuration studied in previous sections. Magnetic activity dominates the uncertainties (in configurations 1, 2, 5, and 6).

However, the OGS signal without magnetic activity, with or without the WGN noise (configurations 3 and 7), also leads to a significant mass uncertainty, again above the objective of 10\% for the 1 M$_{\rm Earth}$ planet. Although the evolution timescales are different, the granulation and supergranulation impact the characterisation of the planet at long periods, here in the habitable zone.
The importance of this contribution was studied in more detail in \cite{meunier19e,meunier20b}. 
Furthermore, when considering a higher level of supergranulation, around 0.7 m/s (red and green stars in Fig.~\ref{contrib}, configuration 1), which seems realistic for the Sun (see Sect.~2.1.2), the uncertainties are slightly increased. For example, the uncertainty for 1 M$_{\rm Earth}$ (PHZ$_{\rm mid}$, $\Delta T_{\rm spot1}$) changes from 56\% to 65\%, and it changes from 30\% to 35\% for 2 M$_{\rm Earth}$. Therefore, although magnetic activity dominates, it will be crucial to improve the correction for all processes, including supergranulation, to reach the objective of 10\%.

Finally, the contribution of the WGN alone is minor at the lowest level, as shown in configuration 4. Triangles and squares indicate the mass uncertainty in this configuration for the higher levels of 0.17 and 0.45 m/s: The uncertainty remains below the 10\% for 0.17 m/s, but it reaches 10 to 20\% for the 0.45 m/s level. Therefore, the contribution of the WGN is important at these levels: HARPS-like RV uncertainties, when obtained with exposure times shorter than one hour, are usually in this range. 
We also note that the impact of the WGN contribution, which is directly related to the signal-to-noise ratio in the spectra, would also be critical for certain correction techniques, especially those considering subsets of spectral lines \cite[e.g.][]{meunier17c,dravins17,dumusque18,cretignier20}.

\subsection{Impact of the yearly gap}

Finally, we compared the performance when using a longer yearly gap, six instead of four months, which may be more realistic for some stars. The comparison was only made for the reference configuration (1), WGN of 0.09 m/s, and is also shown in Fig.~\ref{contrib} as diamonds. The length of the yearly gap contributes slightly, but to a lesser extent than at the level of supergranulation seen above. 
The uncertainty for 1 M$_{\rm Earth}$ (PHZ$_{\rm mid}$, $\Delta T_{\rm spot1}$) changes from 56\% to 60\%, and it changes from 30\% to 33\% for 2 M$_{\rm Earth}$.

\subsection{Tests of other correction methods}
\label{secmeth}

\subsubsection{General approach}

The correction method we applied so far was based on a non-linear relation between RV and $\log R'_{HK}$ and the phase of the cycle (see Sect.~\ref{sec23}). Its objective was to remove most of the signal due to the convective blueshift inhibition in plages, which is the dominating contribution at long periods by far, and which also contributes to the modulation at rotational timescales. The analysis of the residuals performed in Sect.~\ref{sec5} shows that some amount of activity signal remains after the correction, especially at rotational timescales, which is expected to degrade the performance. Here, we test different methods, mostly based on the use of the $\log R'_{HK}$. We consider the  simulations for G2 stars, the reference configuration (1, AR+OGS+WGN, a WGN of 0.09 m/s and the low spot contrast), and a planet of 4 M$_{\rm Earth}$. The rms of the residuals and the performance in terms of mass characterisation are compared with the results obtained in Sect.~3.2 for the same set of simulations, which were obtained based on the model in equation~\ref{eqmodel}.

\begin{figure*}
\includegraphics{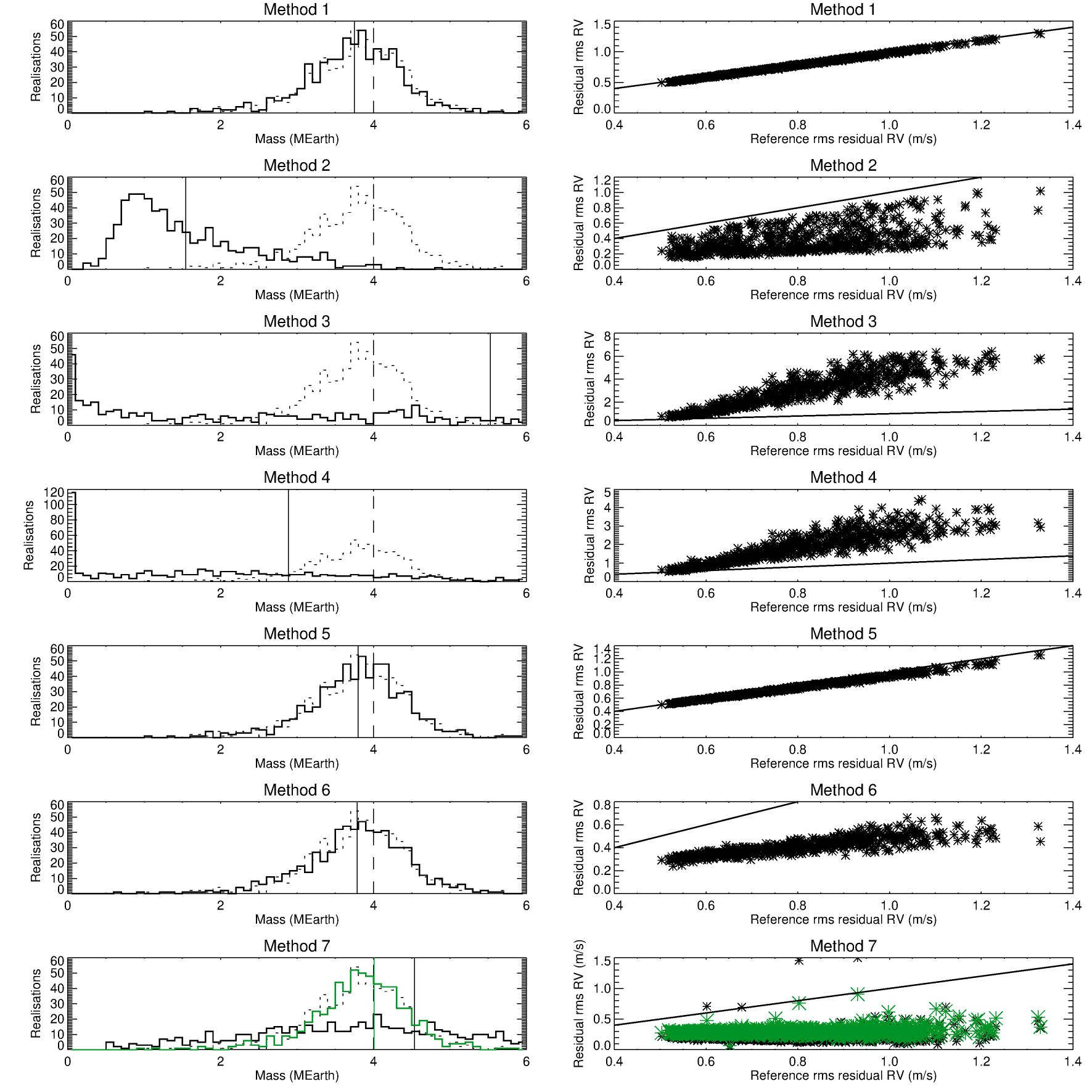}
\caption{
Distribution of the fitted masses (left panel) and rms of the residuals vs. the reference rms (right panels) for a selection of methods (numbers in Table~\ref{tabmeth}).
This corresponds to the follow-up realisations for G2 stars, a planet mass of 4~M$_{\rm Earth}$, the middle of the habitable zone, and a lower spot contrast. The distributions  show the reference mass distribution from Sect.\ref{sec3} (dotted line) and the tested methods (solid lines). For the GP method (7), the results for the analysis performed on the original series are shown as solid black lines, while those corresponding to the analysis performed on the residuals from Sect.~3.2  are shown in green.  The vertical solid lines in the left panels correspond to the average mass estimate for the corresponding distribution. 
}
\label{meth}
\end{figure*}

\subsubsection{Methods}

We present here the different methods we tested. The technical details are given in Sections~\ref{secA1} to \ref{secA6}, and they are summarised in Table~\ref{tabmeth}. Because many variants led to similar results, only a few are illustrated: They are identified by a number in the third column in Table~\ref{tabmeth} and below. We tested five approaches: \\

\begin{itemize}
\item {\it \textup{Denoising at} P$_{\rm rot}$}: 
Denoising based on the presence of peaks in the periodograms of the activity indicators has been used in several studies \cite[][]{boisse11,queloz09}, and also by one of the teams involved in the fitting challenge of \cite{dumusque17}. We are only interested in  planets at long orbital periods. Therefore, we chose to test this type of denoising only at short periods (shorter than 50 days to correspond to the residuals of the rotational modulation), and on the residuals after correction of our reference model. The objective is to determine whether adding this step can help to reduce the dispersion on the fitted masses. The various tests showed that depending on the chosen threshold, the residuals are either to high (weak impact) or far too low (the planetary signal is removed as well). The details are given in Sect.~\ref{secA1}. The example shown in Fig.~\ref{meth} is denoted as example 1. 
\item {\it \textup{Denoising adapted from \cite{rosenthal21}}}: 
\cite{rosenthal21} used a criterion based on the comparison between $\log R'_{HK}$ and RV time series to  attempt to reduce false positives in RV (they did not use it to correct the RV signal). The objective is to identify whether a given RV peak in the periodogram is correlated with the activity indicator. We tested different variations of this principle. The details are given in Sections~\ref{secA2} and \ref{secA3}. The examples shown in Fig.~\ref{meth} are denoted as examples 2 and 3. 
\item {\it \textup{FF' method from \cite{aigrain12}}}: 
These authors proposed the FF' method, basically for simple activity pattern configurations, which is based on the following principle: F is the photometric signal that is  used to correct for the convective blueshift inhibition, and F' is the derivative of the photometric signal that is used to correct for the contrast contribution to RV measurements. Because the convective blueshift inhibition in plages is much better correlated to the $\log R'_{HK}$ than to the photometric signal, we used $\log R'_{HK}$  instead of F. The photometric signal was also produced in the simulations of Paper I and was therefore  used here to obtain F'. The details are given in Sect.~\ref{appC4}. The example shown in Fig.~\ref{meth} is denoted as  example 5. 
\item {\it \textup{Binning}}: 
Since some residuals remain that are associated with stellar variability at low periods, we tested the possibility of binning the data over a typical rotation timescale. This approach was suggested \cite[][]{dumusque11b} to reduce the contribution of oscillations, granulation, and supergranulation. The details are given in Sect.~\ref{appC5}. The example shown in Fig.~\ref{meth} is denoted as example 6. 
\item {\it \textup{Gaussian processes}}:  
GPs have been used by many teams to correct for the rotationally modulated stellar signal in  RV time series \cite[e.g.][]{haywood14,rajpaul15}. 
The groups using this non-parametric technique obtained the best performance in the fitting challenge of \cite{dumusque17}. Even though this is very promising (this method is now widely used), it is also known to possibly overfit the data, so that there is a risk that long-period planets might be absorbed by the GP \cite[][]{langellier21}.  It is therefore important to test this technique on our synthetic time series. Our objective is not to be exhaustive here, however, because many variants exist, but first to test the performance of the correction of the variability of the rotation signal as is usually done in the literature. The details are given in Sect.~\ref{secA6}. The examples shown in Fig.~\ref{meth} are denoted as example 7.
\end{itemize}

\subsubsection{Comparisons between methods}
\label{secA7}

To fully assess the mass estimation performance, the uncertainty on the mass must be considered jointly with the bias on the mass. 
Fig.~\ref{meth}  summarises the main results for a selection of methods from all tests described above. 
Simple denoising methods based on various assumptions on the peaks in the periodograms fall into three categories, as illustrated by the first four methods in the figure. 
For certain methods, the rms of the residuals is much lower than the reference rms, with a mass that is poorly characterised. In other cases, 
the rms of the residuals is much decreased, but the planetary signal has been eliminated by the denoising. In the remaining cases, the rms of the residuals is similar to the reference rms (from Sect.~\ref{sec23}), without an improvement. 
The FF' method (example 5) leads to a small decrease in the rms of the residuals, which may be better if the spot contrast is much higher, but with limited impact on the mass. The binning method leads to a much better rms of the residuals, but without an impact on the mass. Finally, we tested the performance of GPs in two cases (example 7). First, the GP applied to the original time series with a simple long-term correction (two sinusoidal fits), which is similar to what is applied in the literature (long-term trend removed followed by a correction based on a GP at the rotation period). This performs poorly, probably because the long-term signal is not properly removed. When the same GP is applied to the reference residuals (which have a better long-term correction, although not perfect, but part of the rotation signal is also removed), the performance is better, but very similar to the reference correction without GP, even though the rms of the residuals are then very good and close to what would be expected from a contribution of the OGS and WGN alone (which we do not expect the GP to correct for). We insist that a small rms of the RV residuals is   not a guarantee of a good mass estimate because the planet signal can be removed so that it can be associated with a completely biased estimate of the mass, as illustrated with the mass distribution for examples 2-4 in Fig.~\ref{meth}: The average of the mass is very different from the true value. 
Our reference method and examples 1, 5 and 6 leads to a similar bias, with an average estimated mass of 3.8~M$_{\rm Earth}$ instead of 4: The bias is smaller than the uncertainties, however. The GP applied to the residuals computed with our method in Sect.~\ref{sec23} does not improve the mass uncertainty, but appears to improve the bias.

\section{Detection rates and false positives}

\label{sec4}

For the blind search for planets, a first step is to determine whether a planet candidate is detected, and a second step is  to evaluate its orbital parameters and mass.  We first describe the protocol we adopted for these blind tests and then focus on the dependence of the detection results on spectral type and activity level. 

\subsection{Method}

In this new series of blind tests, we again considered the seven spectral types between F6 and K4 and the four masses between 1 and 4 M$_{\rm Earth}$. We focused  on configurations with the low WGN level of 0.09 m/s for the contribution of the photon noise. Because these simulations are much more computationally expensive because the periodograms are computed, 
we considered only 400 realisations for each spectral type and mass. Each set of 400 realisations was  built as follows: for each realisation, a random stellar inclination $i$ was selected, as well as a random time series in the whole data set of simulations for that spectral type. A random variable was used to determine whether a planet is injected, so that half of the 400 realisations have an injected planet on average.
A random period was chosen for the injected planet between PHZ$_{\rm in}$ and PHZ$_{\rm out}$, and its phase was chosen randomly. In addition, due to the stellar inclination and the assumption that the orbital plane is the same as the stellar equatorial plane, the planet mass was multiplied by sin($i$) before injection. The RV time series were then corrected for the activity signal as in the previous section (see equation~\ref{eqmodel} in Sect.~2.3).

Each residual time series was then processed as follows. The Lomb-Scargle periodogram \cite[LSP, ][]{1982ApJ...263..835S} was computed, and a 1\% fap was computed using a classical bootstrap method based on the assumption that the residuals are white noise \cite[as in e.g.][]{dumusque12}.
We chose  to use this approach here because it is classical and fast. A more accurate alternative would be to  directly estimate  the false-alarm level from the distribution of the largest peak of the LSP, as obtained from Monte Carlo simulations over a large set of synthetic time series corresponding to a specific activity regime. This approach was followed in \cite{meunier20} and \cite{meunier20b} for the OGS contribution alone. 
However, in contrast to the OGS contribution, the AR contribution can vary greatly within the same  spectral type, resulting in inhomogeneous regimes with different fap levels. While accurate derivations of fap levels with activity are indeed possible \cite[e.g. using predictive $p$-values, as in][]{sulis22}, these methods are computationally intensive and hence are more suited to  studying  one specific time series  with its own activity regime than to studying thousands of them, as considered here.  The standard WGN-based bootstrap approach for the fap estimation can be dramatically inaccurate for RV detection with stellar activity: This is highlighted in this paper (Sect.~\ref{4.2}), and the detection rate as a function of the true false-alarm rate is also studied (Fig.~\ref{btpdf}). More robust estimates  of the false-alarm level \cite[][]{hara22b,sulis22} are beyond the scope of this study.

The test statistic used in this work is the value of the highest peak in the LSP. If it is above the estimated level in the LSP corresponding to an fap of 1\%, it is considered to be a detection. This classical and fully automated approach is  well adapted to process the large number of synthetic time series.

If a detection is claimed, the next step is to fit the planetary signal  with a sinusoid (the eccentricity is here equal to zero) and estimate its period and mass in this way: The estimated values are then compared to the true planet parameters (if a planet was injected). The study of the distribution of the fitted periods with the true planet periods led us to adopt a maximum difference between the injected and recovered periods of 5\% as a criterion to identify the detected planet. The mass is not used as a criterion here, but some of the true detections will have a better mass estimate than others. We therefore define three categories as follows: 
\begin{itemize}
\item{{\it \textup{Good detection}:} If a planet was injected and the claimed detection is considered to be a valid detection, that is, if the fitted planet period differs from the true period by less than 5\%. }
\item{{\it \textup{Wrong detection}:} If a planet was injected and the claimed detection is not considered to be a valid detection, that is, if the fitted planet period differs from the true period by more  than 5\%. }
\item{{\it \textup{False positive}:} If a planet is detected even though there was no injection, then this is a false positive. }
\end{itemize}
 We separated the notions of wrong detections and false positives to  differentiate the configurations with or without  planet because  the presence of a planet in the signal  leads to a globally different statistical behaviour  compared to the configuration without an injection (e.g. influence of the planet peak sidelobes). 
The number of realisations in each of these three categories is then used to compute three different rates: The good detection rate is the empirical probability that the highest peak of the LSP is found at the true injected period within an  error of at most $5\%$ (i.e. number of good detections divided by the number of realisations with an injected planet), the wrong detection rate (number of wrong detections divided by the number of realisations with an injected planet), and the false positive rate (number of false positives divided by the number of realisations without an injected planet).

We implemented two variants of this protocol:
\begin{itemize}
    \item{Protocol A: The highest peak in the LSP is searched for in the whole period range we considered, that is, 2-2000 days.}
    \item{Protocol B: The highest peak in the LSP is only searched for periods longer than 50 d, assuming that we are only interested here in these  planets, since the periods of the injected planets are in the range 179-1174 days. The idea is to consider that we may separate the search for short- and long-period planets because stellar activity may be dealt with  differently.}
\end{itemize}

\subsection{Dependence on the star spectral type}
\label{4.2}

\begin{figure}
\includegraphics{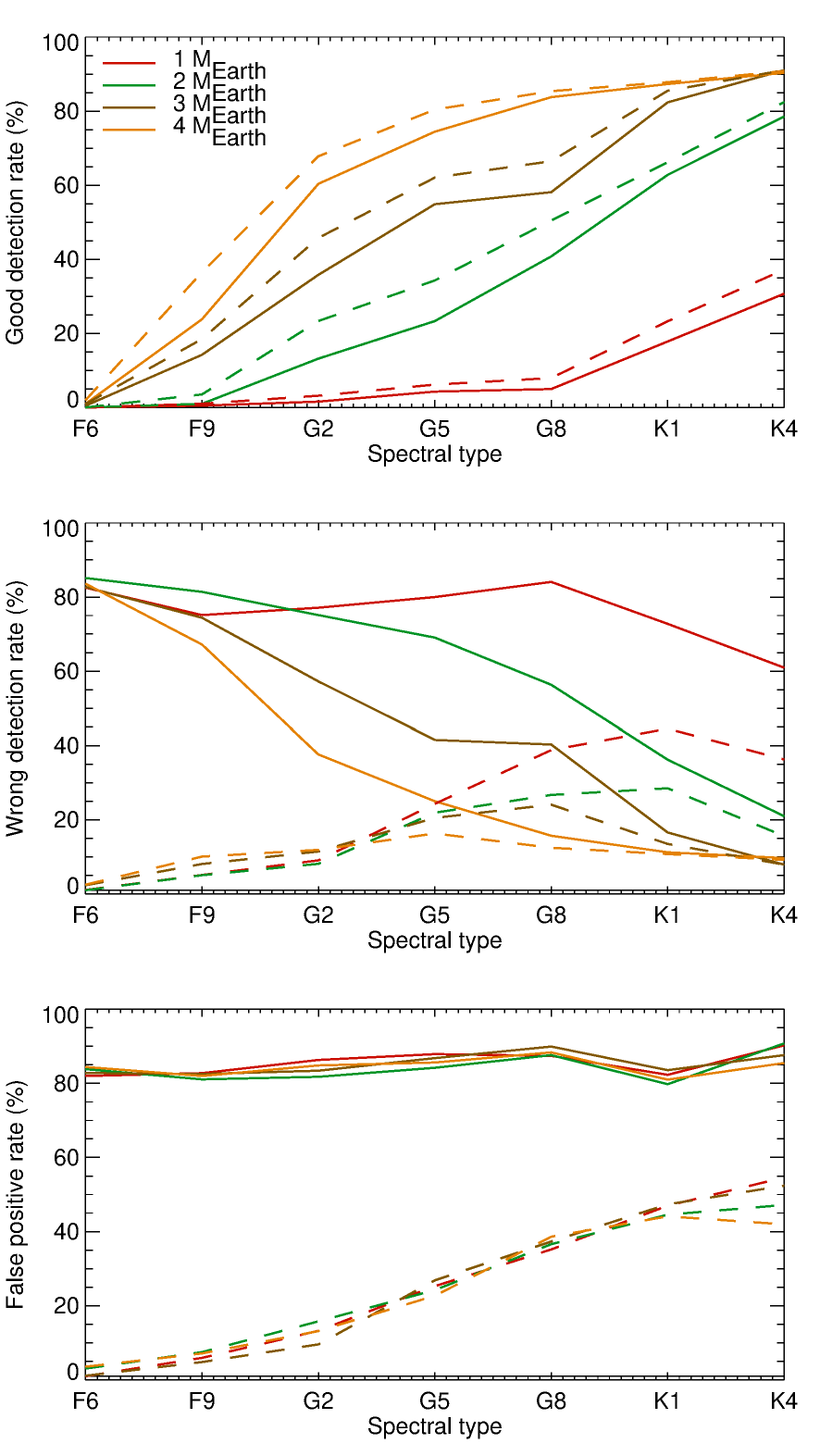}
\caption{
Detection performance vs. spectral type for 1 M$_{\rm Earth}$ (red), 2 M$_{\rm Earth}$ (green), 3 M$_{\rm Earth}$ (brown), and 4 M$_{\rm Earth}$ (orange): Good detection rate (upper panel),  wrong detection rate (middle panel), and false-positive rate (lower panel). Two protocols are tested, A (solid lines) and B (dashed lines). 
}
\label{bt1}
\end{figure}

\begin{figure}
\includegraphics{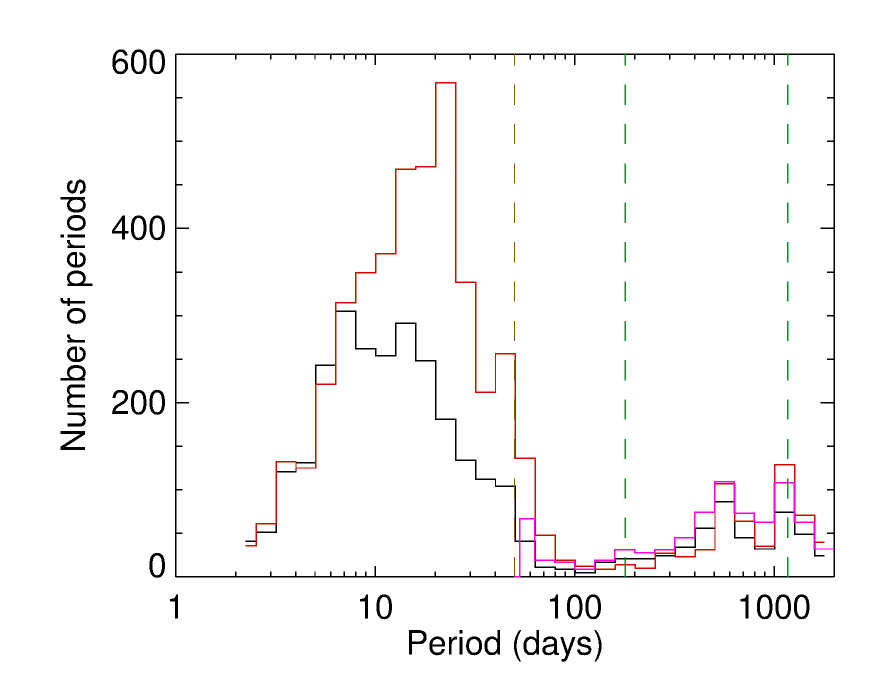}
\caption{
Distribution of the periods of wrong planets (black for protocol A and blue for protocol B) and false positives (red for protocol A and pink for protocol B) for all spectral types and planet masses. The brown  line indicates an upper limit of the rotation periods, and the two   green lines indicate the lower and upper limits of the planetary periods. 
 }
\label{distperfp}
\end{figure}

\begin{figure}
\includegraphics{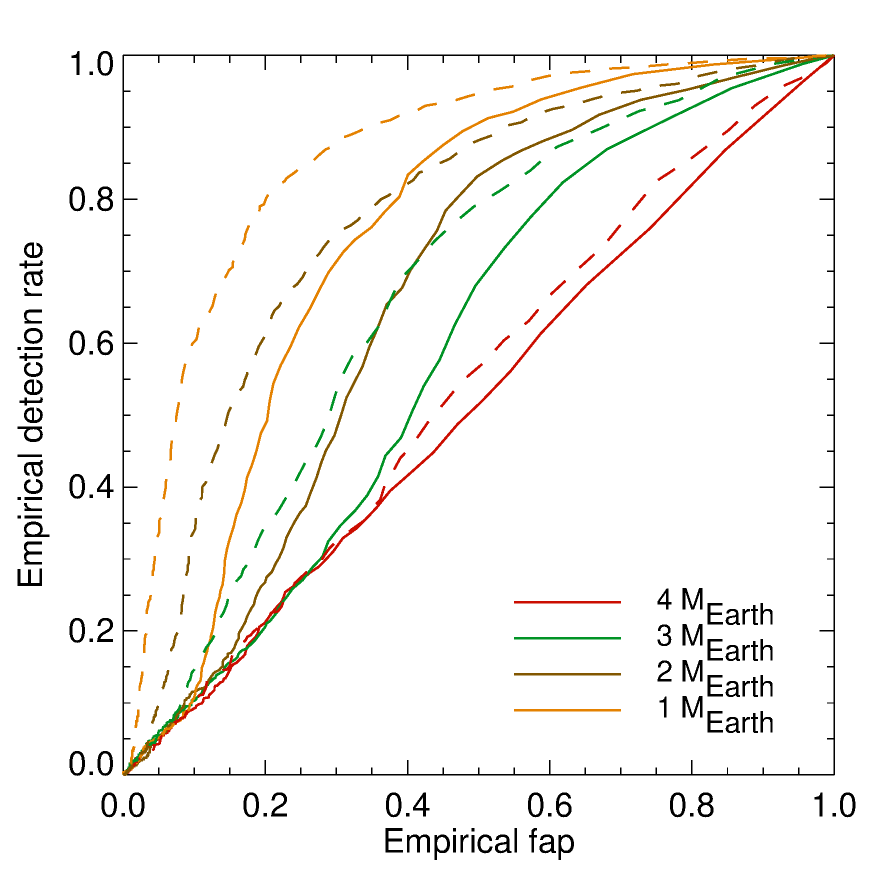}
\caption{
Empirical detection rates vs. empirical fap for all spectral types  for 1 M$_{\rm Earth}$ (red), 2 M$_{\rm Earth}$ (green), 3 M$_{\rm Earth}$ (brown), and 4 M$_{\rm Earth}$ (orange). Two protocols are tested, A (solid lines) and B (dashed lines). 
}
\label{btpdf}
\end{figure}

\begin{figure}
\includegraphics{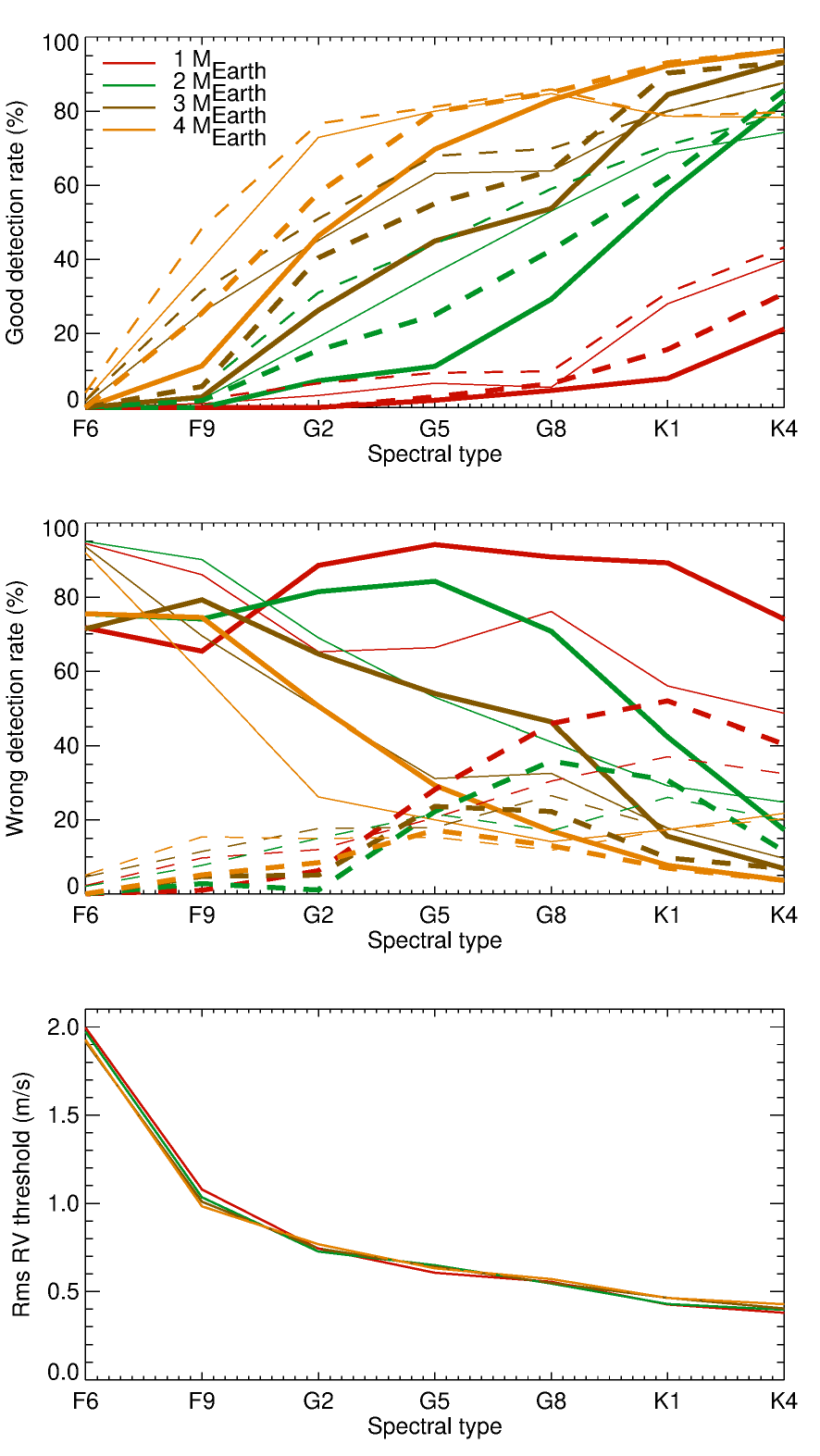}
\caption{
Performance vs.  spectral type for residuals with a high rms (thick lines) and stars with low RV rms residuals (thin lines) for 1 M$_{\rm Earth}$ (red), 2 M$_{\rm Earth}$ (green), 3 M$_{\rm Earth}$ (brown), and 4 M$_{\rm Earth}$ (orange): Good  detection rates (upper panel) and wrong detection  rates (middle panel). The threshold in rms between the two activity classes is the median of the rms values, shown in the lower panel. Two protocols are tested, A (solid lines) and B (dashed lines). 
}
\label{btact}
\end{figure}

 Figure~\ref{bt1} shows 
 the detection performance versus spectral type for the four masses and the two protocols. The good detection rate (top panel) increases towards lower-mass stars for both protocols. This is probably mostly due to the fact that the amplitude of the stellar signal strongly increases towards high-mass stars because the convective blueshift is higher. When using protocol A, the good detection rates are higher than 50\% for the 4 M$_{\rm Earth}$  planet for G and K stars (but the lower panel shows that the actual false-detection rate is more than 20\% in these cases),  and close to zero for 1 M$_{\rm Earth}$  and G2 stars, with a maximum of 30\% for K4 stars (but this requires accepting an even greater false-positive rate). Protocol B leads to slightly higher detection rates, although the trends are very similar.

The wrong planet rate is very high with protocol A, especially for the high-mass stars, and including for a 4  M$_{\rm Earth}$ planet. Protocol B is very efficient in reducing this rate, especially for massive stars because with protocol A, most planets are detected at a  low period, as illustrated in Fig.~\ref{distperfp} (and are consequently wrong detections), typically in the 2-50 d range, which corresponds to rotation periods covered in Paper I, as illustrated in Fig.~\ref{distperfp}. This is expected because the model described in equation~\ref{eqmodel} corrects for the activity signal at both long and short periods, but part of the magnetic activity contribution, namely that due to the spot and plage contrasts, is not removed by this correction because the shape of the RV signal is different from the $\log R'_{HK}$ variability.

When no planet is injected (lower panel in Fig.~\ref{bt1}), the false-positive rate is extremely high with protocol A (around 80\%). It is lower with protocol B, but can still reach high values. 
 Fig.~\ref{distperfp} also shows  false positives at long periods, that is, in the planetary regime, for both protocols: There are still residuals due to activity in the signal, which are studied in detail in Sect.~5. 
That the false-positive rates are far above 1\% means that the residuals do not correspond to a WGN.

A synthetic picture of the achievable trade-offs detection rates versus false-detection rate can be obtained in Fig.~\ref{btpdf} as follows.
 We considered series  without a planet and compute the empirical distribution of the amplitude of the highest LSP peak. We did the same for time series with a planet injected. For each of those two distributions, we varied a threshold, $\gamma$, and counted the fraction of values above it. This allowed us to plot an empirical  detection rate that includes both good and wrong detections  (because it corresponds to the highest peak in all realisations)  and a false-positive rate. The results are shown in Fig.~\ref{btpdf} for all spectral types, and they are shown separately for the different masses. 
A curve along the diagonal, as is observed for 1  M$_{\rm Earth}$, means that the test is inefficient, that is, it does not allow us to distinguish between a time series with an injected planet and a time series without a planet, given the properties of the RV residuals and the correction method. This is consistent with the results shown in Fig.~\ref{bt1}.  The performance is better with increasing mass, although even then there is a threshold in fap below which the test is not efficient.
These results depend on the method we used to correct for the stellar signal, and they are therefore valid for the model described in equation~\ref{eqmodel}. The models tested in Appendix~\ref{appCC} do not exhibit a much better performance, however, so that these results are probably valid in a broader context.

In addition, the protocols presented here considered only the highest peak. Protocol A is based on the widest range in periods, and as a consequence, it gives a higher rate of wrong planets or false positives. However, the highest peak, when above the considered fap threshold, is not always the only peak that satisfies this criterion. We therefore also examined the other peaks above the fap threshold for the realisations with a wrong planet detection. In some cases, the true planet peak was also present and above the fap threshold (although not the highest peak), and therefore, this particular peak is not retrieved as a detection in our analysis. 
This is particularly true for the low-mass stars. This is consistent with the results of protocol B.

\subsection{Dependence on the star activity level}
\label{sect43}

\begin{figure}
\includegraphics{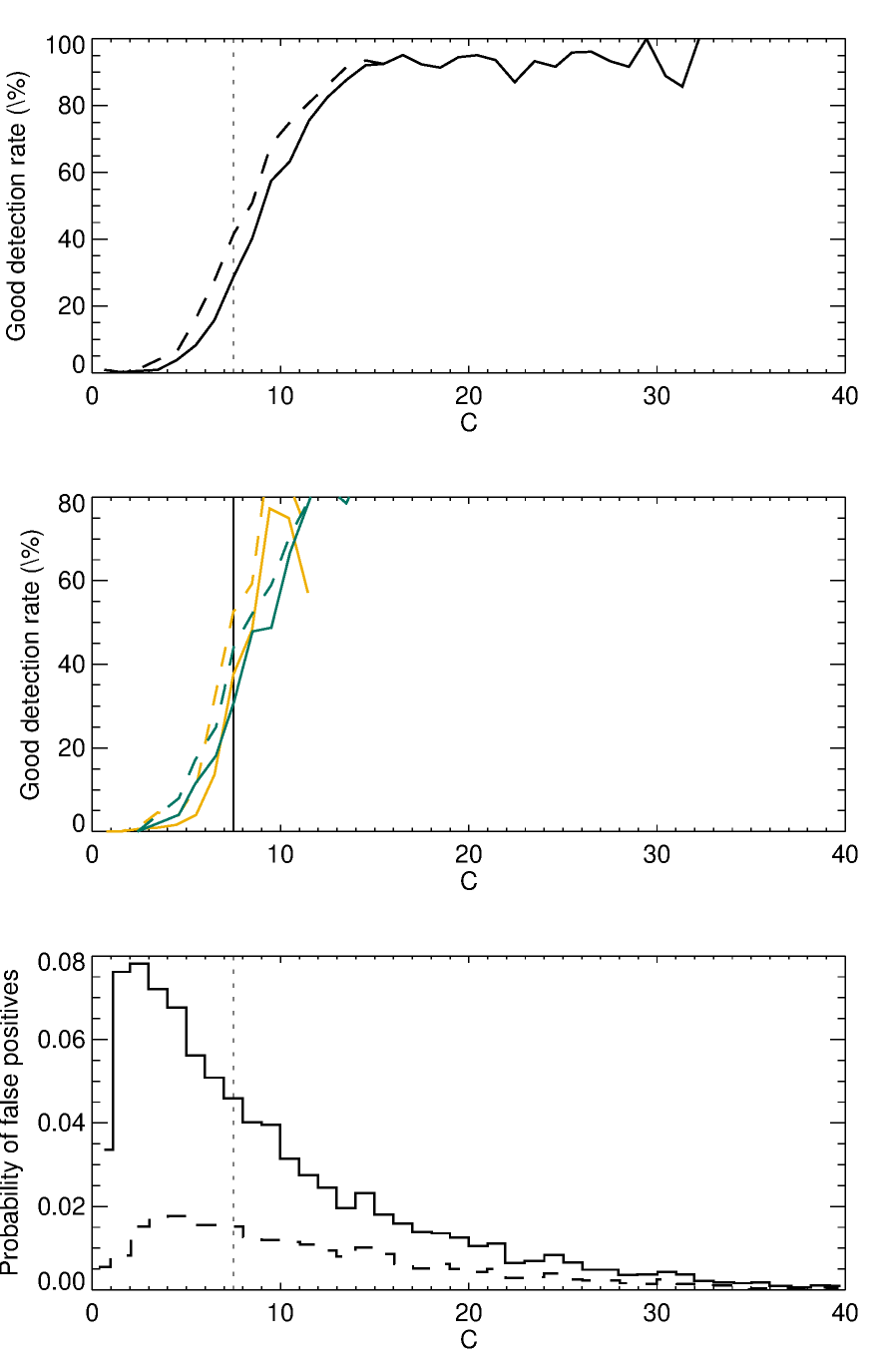}
\caption{
Good detection rates vs. criterion C defined in \cite{dumusque17}, see Eq.~\ref{eqc} for a definition, for all planet masses and all spectral types (upper panel) for protocol A (solid line) and protocol B (dashed line). The vertical line indicates the 7.5 level from \cite{dumusque17}. The middle panel shows the same curves for F9 stars (yellow) and K4 stars (blue). The lower panel shows the probability (computed as the number of false positives in that bin divided by the total number of false positives) of false positives vs. $\widehat{C}$, a value of $C$ estimated from the fitted mass for time series that have no planet (in contrast to the two upper panels) and that created a false positive, for protocol A (solid line) and protocol B (dashed line).
}
\label{coeffc_bt}
\end{figure}

In this section, we investigate how the stellar activity level impacts the detection performance for each spectral type. For a given spectral type, we  expect that more active stars  should lead to time series with higher residuals after correction for activity. We split the time series of a given spectral type into two activity classes (lower and higher),
those with a low rms of the RV residuals, and those with a high rms. The threshold was chosen to be  the median of the RV rms over the considered sets. The detection rates 
are shown in Fig.~\ref{btact}, as well as the threshold between the two sets of realisations, which is for example 0.8 m/s for G2 stars.  The different curves show the same trends globally, so that the wrong detection rate remains high even for low RV rms residuals for both protocols. There is a difference of about  5-20 \%  between the two domains of rms, however, with better good detection rates and lower wrong  detection rates for the low rms residuals.

Another way to study this dependence is again to use criterion C from \cite{dumusque17}, defined in Eq.~\ref{eqc}. As before, we cover here a wide range of RV rms of the residuals, and 1-4 M$_{\rm Earth}$, but only one value for the number of observations. We  considered all spectral types and planet masses, and computed the good planet detection rate in each bin in C\dm{.} 
This rate is shown in Fig.~\ref{coeffc_bt} (upper panel) for the two protocols. The threshold of 7.5 from \cite{dumusque17} corresponds to a good detection rate of about 30\%. On average, a value of C around 10 would be necessary to reach 50\%, and a value of 14 is required to reach 80\%. When considering the spectral types separately, there is a small dispersion, but no obvious trend. This result appears to be robust with spectral type, as shown in the middle  panel of Fig.~\ref{coeffc_bt}. As an example, a good detection rate of 75\% (C$\sim$9) for F9 stars would be reached for a 3.6 Earth mass planet in the middle of the habitable zone. 
This is indicative because, as seen before, the level of false positives remains high, as illustrated in Fig.~\ref{coeffc_bt} and in the lower panel of Fig.~\ref{bt1}. 
As an example, we considered one of the false positives that was detected when no planet was injected, with a fitted mass of 1~M$_{\rm Earth}$, for example. For any time series, we can estimate a posteriori a value of $C$, for instance  $\widehat{C}$, corresponding to this planetary mass and the rms of the residuals for that time series, for example 10. 
Based on the top panel of Fig.~\ref{coeffc_bt}, we might conclude that in this example, this is a safe regime and that the detection is likely to be robust, which is not the case. When we consider all false positives we obtained out of the realisations without an injected planet (all spectral types and 1~M$_{\rm Earth}$), the  $\widehat{C}$ values of a substantial fraction are higher than 10 (34\% of all false positives for protocol A and 47\% for protocol B), as shown in the lower panel of Fig.~\ref{coeffc_bt}. Summing all bins in this panel provides the total fraction of false positives (85\% for protocol A and 25\% for protocol B), which typically corresponds to the average level in the lower panel of Fig.~\ref{bt1}. 

We finally note that  the curves shown in  all panels of Fig~\ref{coeffc_bt} depend on the threshold 
that is chosen to tune the fap: A lower target fap would have provided lower good detection rates, and as a consequence, the same value of C=7.5 would have corresponded to a much lower detection rate.  In addition, C does not account for other aspects that strongly impact the detection rate, such as the temporal sampling or the coverage of the observations. These results show that a blind use of the C criterion as a rule of thumb to evaluate  detection performance can be extremely hazardous.

\section{Analysis of RV residuals}
\label{sec5}

As we showed in the previous section, the wrong planet rates (planet injected) and false-positive rates (no planet injected) are very high and suggest  that the residuals are not white noise because the false-positive rate is above the 1\% required fap level.
The objective in this section is therefore to study the residuals obtained in the previous section after applying the model in equation~\ref{eqmodel}. We focus on planets located in the HZ, or closer in. We wish to evaluate what remains to be corrected for to reach better performance. The analysis was done in the period range 2-2000 days to cover rotation shorter than 50 days, and the planetary range. 
For this purpose, we analysed the maximum power in LSP in different period ranges described in this section. In addition, the rms of the residuals after binning to extract a long-term remaining variability is discussed in Appendix~\ref{appBB}.
These analyses can in principle be applied to the residuals from the follow-up blind tests (Sect.~3, either before or after planet removal) or to those from the detection blind tests\footnote{While this iterative subtraction-based approach for activity correction is most popular, we note that  estimates of the activity signal can also be used to directly calibrate the periodogram \cite[][]{sulis22}, which may perform better.} (Sect.~4, before planet removal because the planet is often only poorly identified, however).

\begin{figure}
\includegraphics{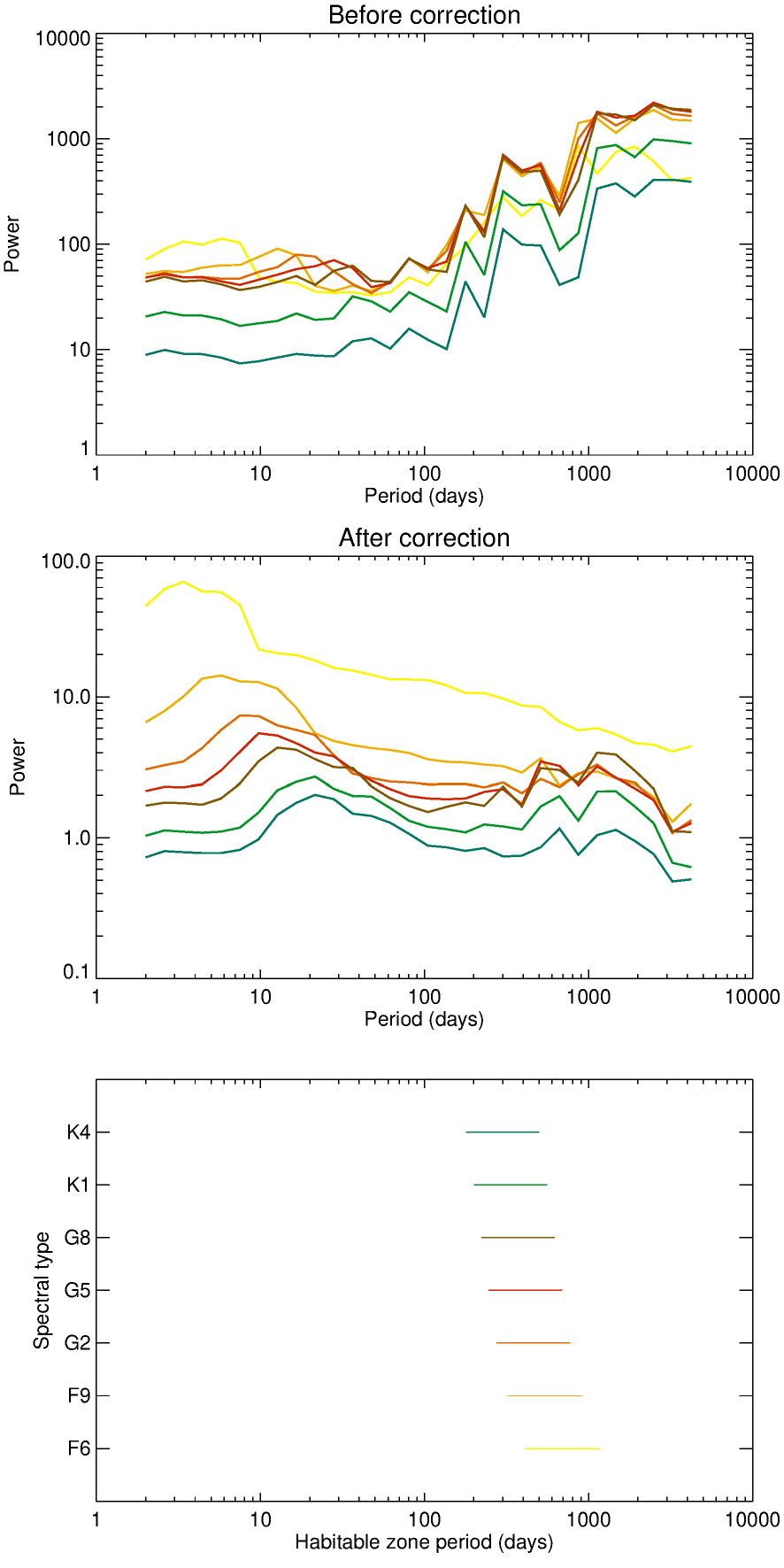}
\caption{
Maximum power in the LSP, averaged over the realisations, vs. period range for the seven spectral types before (upper panel) and after correction (middle  panel). The residuals are from  the follow-up analysis for $\Delta T_{\rm spot1}$, the middle of the habitable zone, and 1~M$_{\rm Earth}$ (fitted planetary signal removed). The habitable zone for each spectral type is  shown in the lower panel, which also indicates the colour code for each of them (from top to bottom in this panel: K4 to F6).  
}
\label{residupow2}
\end{figure}

There are more realisations in the follow-up blind tests. We therefore  show results from the residuals from Sect.~3 after correction for the activity and planetary signal. We then  computed the LSP for each residual for periods between 2 and 6000 days. For each realisation, we  computed the maximum power in different bins in period (we used 30 bins of equal size in $\log(P)$). We used bins in order to account for the fact that the stellar signal corresponds to different periods depending on the realisation 
and exhibits many peaks. This is due for example to  differential rotation, to the limited lifetime of the structures, and to the irregular observation window. 
This power was compared before and after correction to identify the decrease in power and what remains. The maximum power in a given period bin was first averaged for all realisations (low spot contrast, planet in the middle of the habitable zone) of a given spectral type. The results are shown in Fig.~\ref{residupow2}.

The gain brought by the considered AR activity model is strong at all periods, both at short and long timescales. The power at long periods dominates before correction, while the  power at short periods dominates after correction for all spectral types. This is consistent with the fact that most false positives and wrong planets have short periods. 
The AR model does not take the contribution from the spot and plage contrast into account, which also corresponds to the rotation periods (see Sect.~\ref{appC4} for a discussion) as well as the OGS and WGN contributions. We also observe a bump in power close to the habitable zone, ranging from a few hundred days to a few thousand days,
 which is consistent with a fraction of the false positives or wrong planets at these periods (see Fig.~\ref{distperfp}).

\begin{figure}
\includegraphics[width=9.5cm]{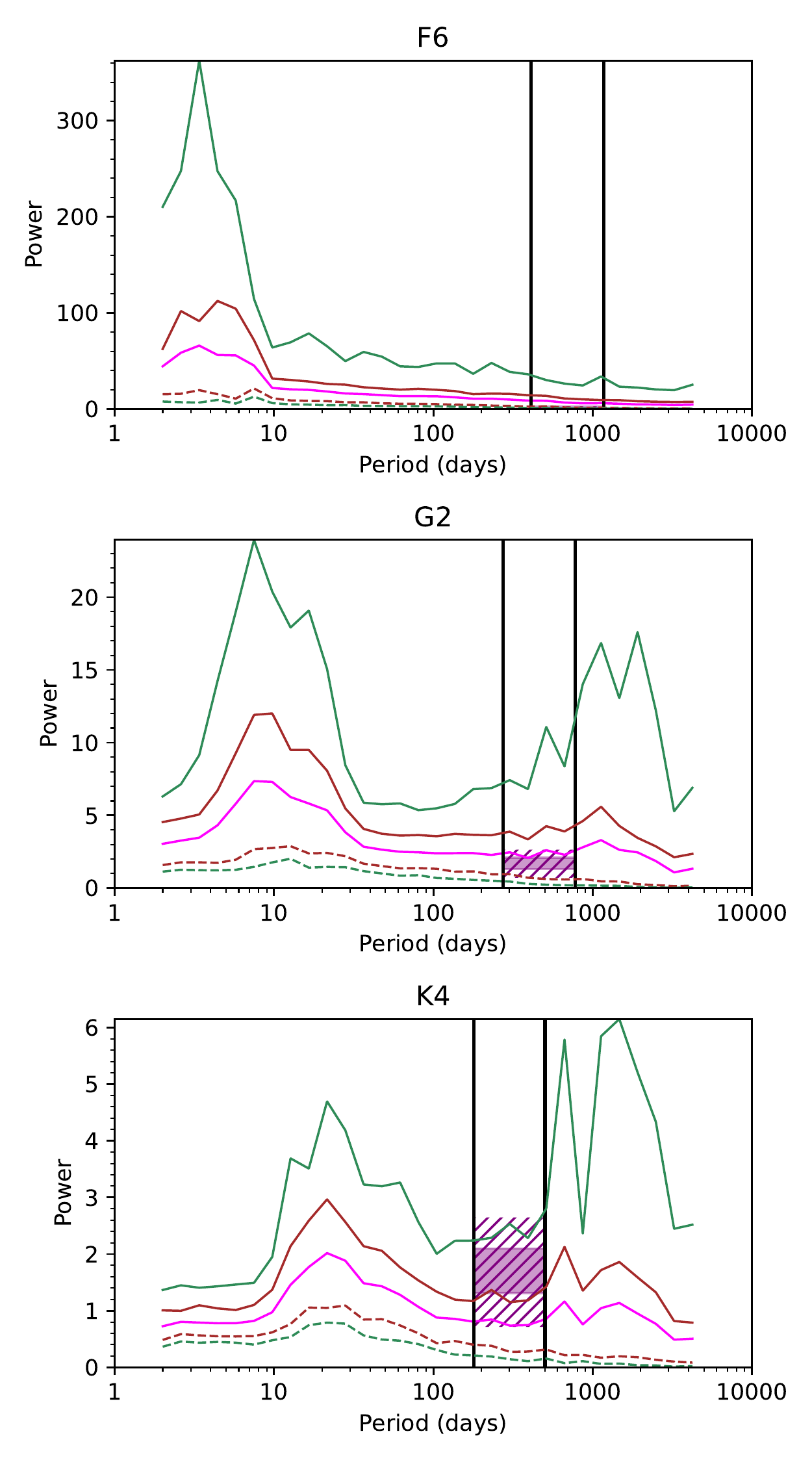}
\caption{
Envelope of the different LSP realisations after correction at $\pm 1 \sigma$ (brown) and $\pm 2 \sigma$ (green). We show the  upper envelope  (solid line) and  lower envelope  (dashed line) for the F6, G2, and K4 spectral types (from top to bottom).  
The average power is  plotted in pink.
The typical power due to a 1~M$_{\rm Earth}$ planet alone (without activity) in the habitable zone (delimited by the two vertical lines) is shown in purple. The filled rectangle corresponds to the $\pm 1 \sigma$ levels, and the dashed lines indicate the full extent of possible values for the planet signal. 
}
\label{residupow3}
\end{figure}

The average curves shown in Fig.~\ref{residupow2} hide the fact that the residuals are highly diverse. Figure~\ref{residupow3} shows the range covered by individual LSP before averaging over the 648 realisations for G2 stars. The realisations are highly dispersed. The purple rectangle gives an idea of the expected power for a 1M$_{\rm Earth}$ planet alone (no stellar contribution) in the habitable zone for a comparison with the residuals: Many LSP show a power above this reference level, explaining the very poor detection rate for these planets around G-type stars obtained in Sect.~4. We conclude that better correction techniques need to improve the performance at all timescales, with better models, and mostly in the range of rotational modulation. Better models at long periods, which typically correspond to a fraction of the cycle period, are also necessary in the habitable zone and at slightly longer periods.

\section{Conclusion}

We performed blind tests to characterise the impact of stellar variability on the mass of Earth analogues in RV follow-ups with HARPS-like instruments (and possibly ESPRESSO) and blind searches. This statistical study combined important properties, most of them never combined before. The properties are 1/  very large set of physics-based synthetic time series; 2/ most physical processes (oscillations, granulation, supergranulation, spot and plage contrasts, and inhibition of the convective blueshift in plages); 3/ variability at all timescales (short, rotation modulation, and cycle timescales);  4/ injection of planets in various configurations (both for RV follow-up and blind search) and retrieval of the signal; and finally, 5/ characterisation of the RV residuals after correction. 
We obtain the following main results after a correction for stellar variability based on a non-linear relation with $\log R'_{HK}$ and cycle phase, assuming 1000 nights of observations (one-hour exposure time each) spread over ten years and a one-planet system: 
\begin{itemize}
\item{The uncertainties on the masses estimated with the RV method performed as follow-up of transit observations reach 40\% for a 1~M$_{\rm Earth}$ planet orbiting in the habitable zone of a G2 star. This is  significantly above the 10\% objective for PLATO, despite a very good temporal sampling (Sect.~3). The 10\% objective is only reached for a 3~M$_{\rm Earth}$  or for lower-mass stars (K4). }
\item{The performance is not improved by using a variant of the FF' method ( \cite{aigrain12}), Gaussian processes correcting for the modulation at the rotation period nor averaging (Appendix~\ref{appCC}). We note that Gaussian processes performed without fitting the planetary signal at the same time causes the planetary signal to become absorbed by the GP, even with a training of the GP hyperparameters on the $\log R'_{HK}$ before fitting the RV time series, meaning that this technique cannot be applied to improve the correction in the detection blind tests. Simple denoising techniques should be avoided. We also provide an estimate of the residual RV rms that should be reached to obtain 10\%.  }
\item{We finally compared criterion C defined in Eq.~\ref{eqc} and proposed by \cite{dumusque17}, with the threshold of $\sim$7.5 for a good performance in their blind test. This threshold corresponds here to a mass uncertainty of typically 20-30\%, while a value around 10 would be better for a 10\% uncertainty. We finally provide the maximum rms of the residuals as a function of spectral type that would be necessary to reach this objective.}
\item{Blind tests performed to quantify planet  detectability perform very poorly (Sect.~4), especially for the more massive stars, with very low good detection rates and very high levels of false positives and wrong planet rates (much higher than 1\%). Our results are linked to the test we used (maximum of the LSP), which is very widely used, but the main issue remains the activity model, which does not entirely correct for the stellar contribution. 
A value of 7.5 for the C criterion corresponds to detection rates of about 30\%, but with a high level of false positives, mostly at low periods. A significant fraction of these false positives are at long periods, however. 
}
\item{ A bootstrap analysis relying on the assumption that the residuals are white leads to extremely underestimated fap levels when the residuals contain residual activity, which is the case of current activity correction models (Sect.~4). Alternative robust methods must be used  to provide reliable false-alarm rates \cite[][]{sulis22,hara22,hara22b}.}
\item{The analysis of the residuals shows that it will be necessary to improve the residuals at both short and long timescales (typically about a fraction of the cycle period) (Sect.~5). In addition, even if magnetic activity dominates, it will be necessary to take the other contributions (e.g. supergranulation) into account to reach residuals that are good enough to obtain a 10\% precision on the mass.   }
\end{itemize}

We recall that the main remaining stellar process, meridional circulation, is not included in these simulations. We also note that we assumed a very good instrumental behaviour (e.g. no temporal variation of the instrumental noise).

Finally, we note that to produce a very large set of time series, the RVs were computed analytically;  therefore, we cannot test sophisticated techniques that would rely on additional activity indicators from the CCF (cross-correlation function) or that are based on the CCF shape \cite[e.g.][]{collier21}, on the use of different spectral lines \cite[][]{meunier17c,dravins17,dumusque18,cretignier20}, or on the use of full information about the spectra \cite{cretignier22}. However, the main method used in this paper still provides a very useful insight into the current  limitations due to stellar activity, the objective of reaching a better performance, and  the performance gap that remains to be filled.

\begin{acknowledgements}

This work was supported by the "Programme National de Physique Stellaire" (PNPS) of CNRS/INSU co-funded by CEA and CNES.
This work was supported by the Programme National de Plan\'etologie (PNP) of CNRS/INSU, co-funded by CNES.

\end{acknowledgements}

\bibliographystyle{aa}
\bibliography{biblio}

\begin{appendix}

\section{Correlation between the mass uncertainty and residual RV rms}
\label{appAA}

\begin{figure*}[h]
     \centering
    \subfigure{%
       \includegraphics[width=0.9\textwidth]{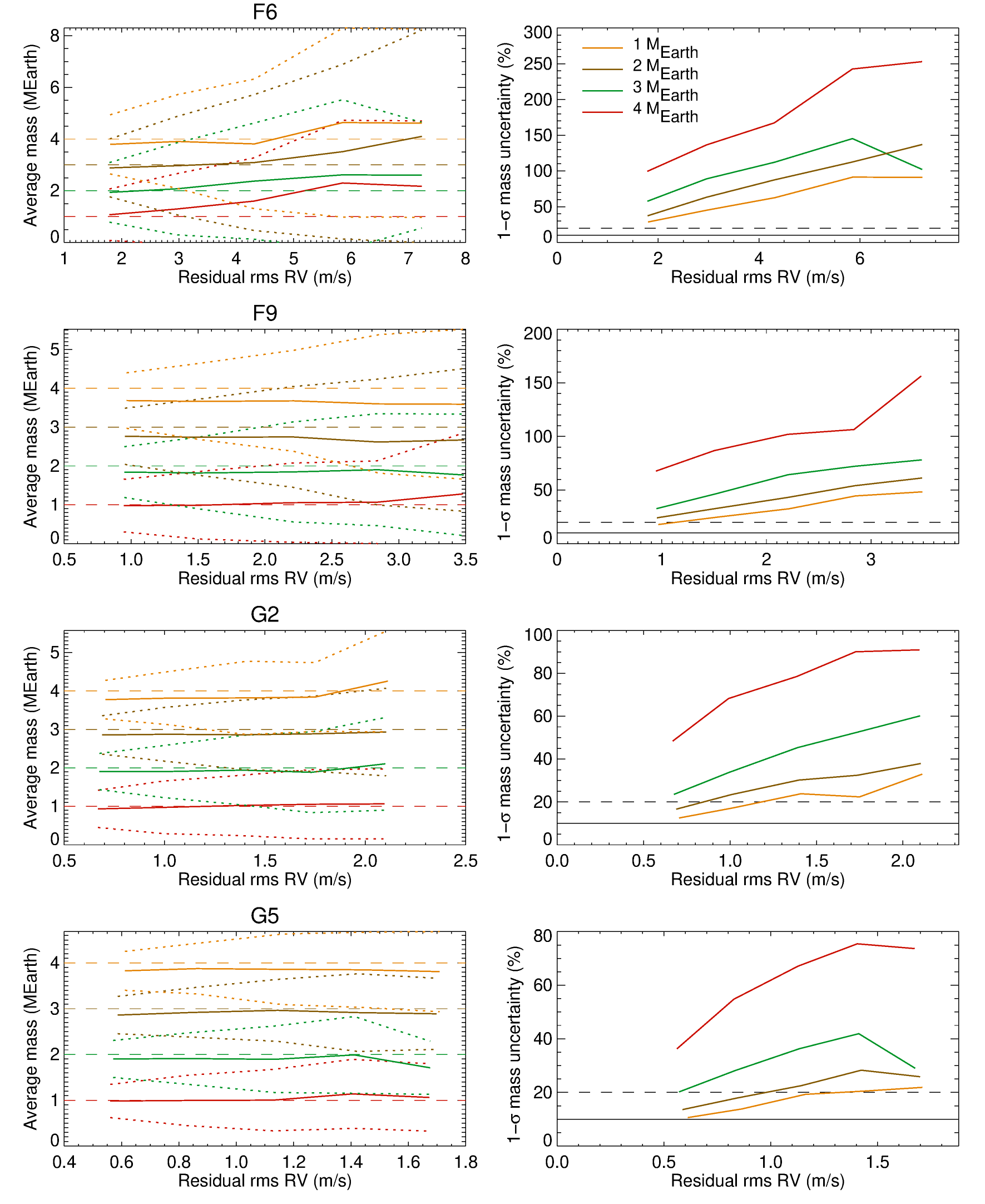}}%
    \caption{Average mass (left panels) and 1$\sigma$ mass uncertainty (right panels) vs. residual of the RV rms after a correction based on the model in equation~\ref{eqmodel} for the different spectral types (one per panel) and masses: 1 M$_{\rm Earth}$ (red), 2 M$_{\rm Earth}$ (green), 3 M$_{\rm Earth}$ (brown), and 4 M$_{\rm Earth}$ (orange). In the right panels, the  solid horizontal line indicates the 10\% objective for PLATO, and the dashed line shows an indicative level of 20\%. The dotted lines in the left panels indicate the $\pm$1$\sigma$ uncertainty envelope. For each row, the results are averaged over all configurations of habitable-zone periods and spot contrasts.      \label{niveau1}} 
  \end{figure*}
  \addtocounter{figure}{-1}
  \begin{figure*}[h]
     \centering
    \subfigure{%
       \includegraphics[width=0.9\textwidth]{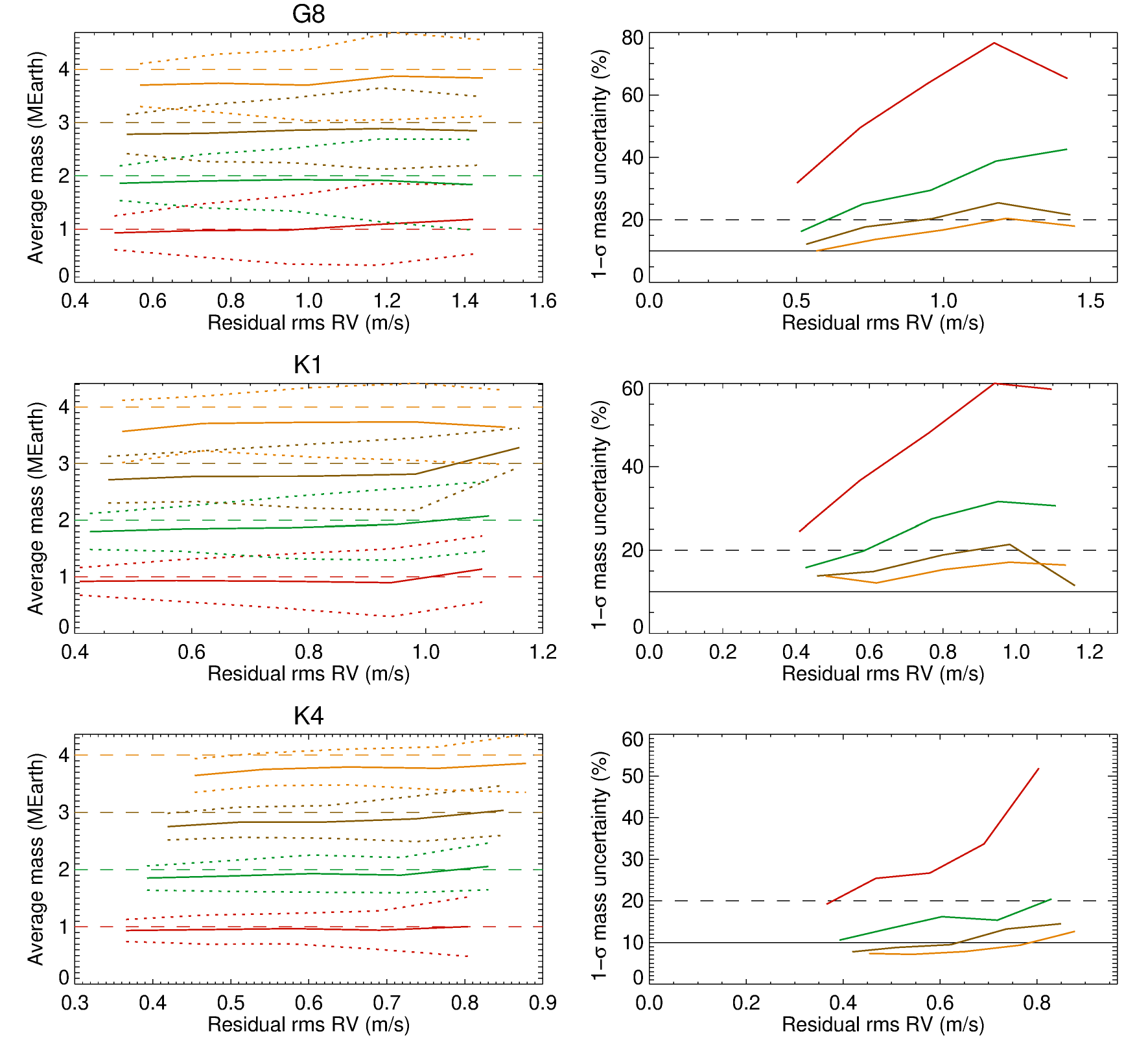}}%
    \caption{Average mass and 1$\sigma$ mass uncertainty (continued).}%
      \end{figure*}


\begin{figure}[h]
\includegraphics{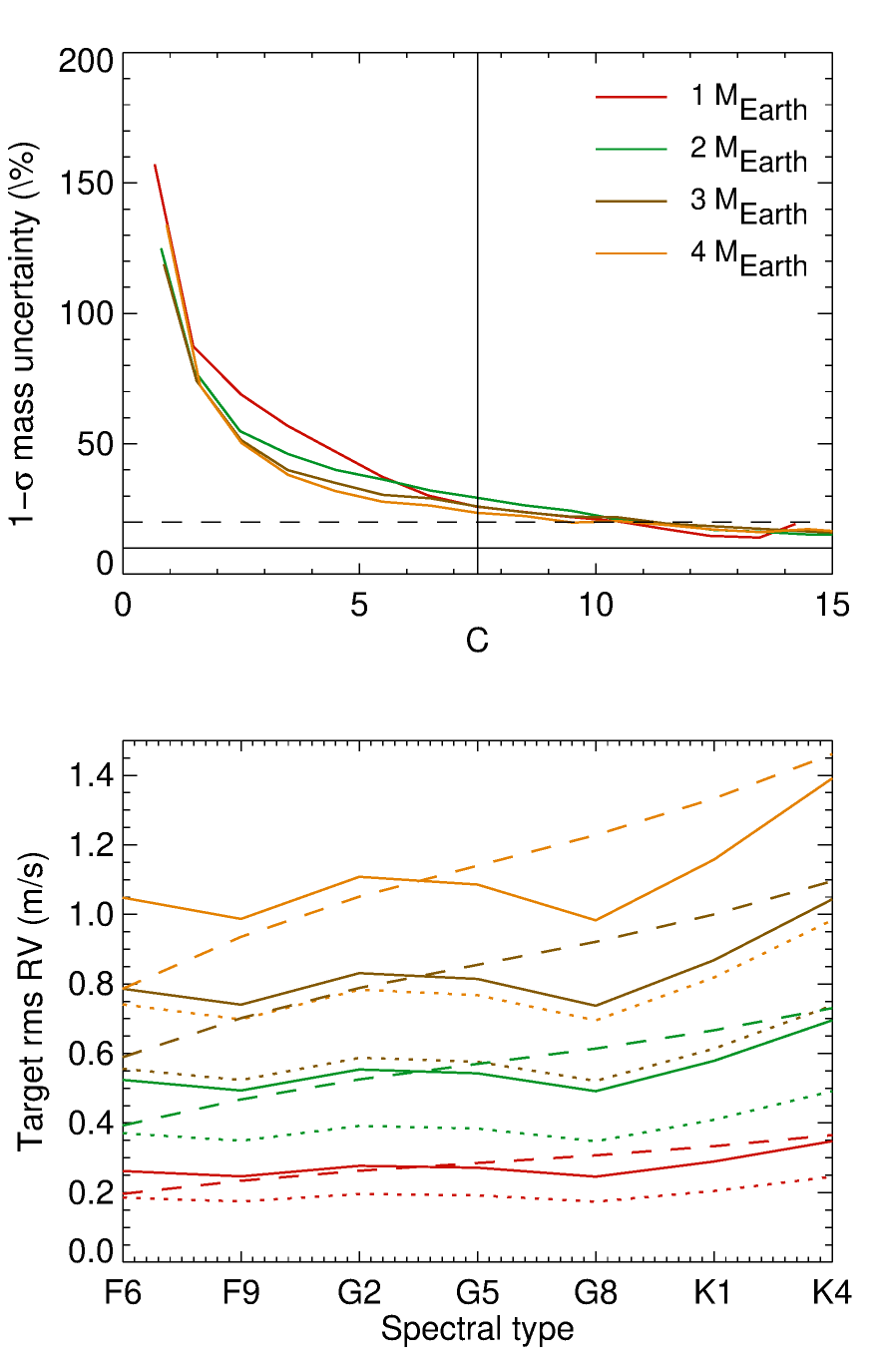}
\caption{
Mass uncertainty (upper panel) in the follow-up blind tests vs. the criterion defined in \cite{dumusque17}; see Eq.~\ref{eqc} for the definition. We show the four masses: 1 M$_{\rm Earth}$ (red), 2 M$_{\rm Earth}$ (green), 3 M$_{\rm Earth}$ (brown), and 4 M$_{\rm Earth}$ (orange). The 10\% and 20\% uncertainty levels are indicated.   
The lower panel indicates the limit in RV rms corresponding to a mass uncertainty of 20\% for the four masses with three assumptions according to this criterion C: 1000 points and limit C depending on spectral type (solid lines), 1000 points and limit C equal to 10 for all spectral types (dashed lines), and 500 points and limit C depending on spectral type (dotted line). 
}
\label{coeffc_fu}
\end{figure}

We focus here on the estimation performance of the mass in terms of bias (i.e. the average value of the estimated mass minus the true value) and dispersion (i.e. the rms of the estimated mass values, referred to below as  1$\sigma$ mass uncertainty) as a function of the activity level (rms of the residuals). We recall that this analysis corresponds to 1000 nights of observations over ten years. 
We considered the residuals after removing the activity model and the fitted planet signal. We then considered different bins in rms. Figure~\ref{niveau1} shows the average mass (left panels) and the 1$\sigma$ mass uncertainty (right panels) as a function of the RV rms of these residuals for the different spectral types (one per panel) and the four masses. The three positions in the  habitable zone  and the two spot contrasts are considered. There is a bias on the mass, especially for large uncertainties, but it is lower than the uncertainty. The mass uncertainty clearly decreases when the rms is lower. The lowest rms obtained here usually does not allow us to reach the 10\% for 1 M$_{\rm Earth}$ for example, except for the K4 stars. However, it is possible to estimate an approximate rms threshold that would allow us to reach this 10\% level. For example, for a 1 M$_{\rm Earth}$ planet, we estimate that a residual at the 0.3-0.35 m/s level would allow us to reach the 10\% level for G and K stars alone. The threshold is about 0.6 m/s for a 2 M$_{\rm Earth}$. 
The left panels in Fig.~\ref{niveau1}  show the average fitted mass versus the RV rms of the residuals. There is a trend in the bias on the estimated mass in a few cases, mostly for low-mass planets and high stellar masses, but it is always below the 1$\sigma$ uncertainties.

In the blind test presented  in \cite{dumusque17}, it was found that criterion C defined in Eq.~\ref{eqc} constituted a useful test of the quality of the planet recovery: the results were globally good for C$<$7.5, while they corresponded to a poor performance for C$>$7.5 in general.
For comparison, we represent the mass uncertainty as a function of this criterion in Fig.~\ref{coeffc_fu}. The curves are similar for the four masses, and the 7.5 threshold corresponds to an uncertainty on the mass slightly above 20\%. The objective of 10\% is then closer to a value of C$\sim$10. The highest C values depend on the planet mass because C increases with K$_{\rm pla}$ and therefore with the mass.  When the different spectral types are considered separately, there is a small dispersion, so that the threshold is in the range 8-12. This therefore also gives a good criterion in terms of performance to reach to be able to obtain a 10\% mass uncertainty. As before, $N_{\rm obs}$ is always the same here (1000), but the rms covers a wide range, which is realistic. 
For practical purposes, this limit in C can be converted into a limit in rms of the residuals as follows. For each spectral type, the requested value of C corresponding to 10\% was used to compute the RV rms for the four planet masses and 1000 points. These RV rms, which are indicative of the objective of 10\%, are shown in the lower panel of Fig.~\ref{coeffc_fu} as a solid line. The dashed line corresponds to the limit of 10 in C for all spectral types. The dotted line corresponds to 500 points instead of 1000.
We note that for the lowest mass (1 M$_{\rm Earth}$), these rms are very low and below the typical value for the OGS and WGN contribution for G2 stars for example. This means that it is not sufficient to consider the AR contribution to reach these low levels. 
In conclusion, we underline that the use of the C criterion must be taken with care. This criterion assumes a single sinusoid in perfectly white noise sampled on a regular  grid without a gap. These  conditions are not met in RV observations. In practice, the value of C that must be reached to guarantee a specific performance (mass estimation for instance) is  highly variable and  depends on various factors  (spectral type, nature and level of stellar activity, and time sampling). In the context of the present blind test, in which the detection was made independently (via the transits), the criterion can be of interest, but its limitations must be kept in mind (see Sect.~\ref{sect43}).


\section{Tests of other correction methods}
\label{appCC}

In this appendix, we detail the different correction methods 
that we discussed in Sect.~\ref{secmeth}, mostly based on the use of the $\log R'_{HK}$, in follow-up blind tests as in Sect.~\ref{sec23}. We consider the  simulations for G2 stars, the reference configuration (example 1, AR+OGS+WGN, a WGN of 0.09 m/s and the low spot contrast), and a planet of 4 M$_{\rm Earth}$.
Sections~\ref{secA1} to \ref{secA6} present the results for six methods (summarised in Table~\ref{tabmeth}). 
Because many variants led to similar results, only a few are illustrated. They are identified by a number in the third column in Table~\ref{tabmeth}. \\

\begin{table*}[h]
\caption{Tested methods in addition to the protocol described in Sect.~\ref{sec23}. }
\label{tabmeth}
\begin{center}
\renewcommand{\footnoterule}{}  
\begin{tabular}{lll}
\hline
Method & Variants & Number \\ 
   &   & in figures \\ \hline
   Model based on $\log R'_{HK}$ and cycle  &  &  Reference\\
   \hspace{0.5cm} phase (Eq.~\ref{eqmodel}) & & \\
Denoising Prot & Ca peak above 10\% maximum power &    \\
 & Ca peak above noise  &  1 \\
 & Ca peaks above 1\% level & 2    \\
 & RV peaks above 1\% level &     \\
 & RV peaks above 1\% level in residuals &     \\
Denoising adapted  & Fixed threshold of 0.9 & 3  \\
\hspace{0.5cm} from \cite{rosenthal21} & Variable threshold &   \\
Denoising based on correlation  & Thresh. 0.8, all P, Phase RV &  \\
\hspace{0.5cm} at long periods   & Thresh. 0.8, elimination Ppla, Phase RV &  \\
   & Thresh. 0.8, all P, Phase $\log R'_{HK}$ &  \\
   & Thresh. 0.8, elimination Ppla, Phase $\log R'_{HK}$ &  \\
   & Thresh. 0.9, all P, Phase RV & 4  \\
   & Thresh. 0.9, elimination Ppla, Phase RV &  \\
   & Thresh. 0.9, all P, Phase $\log R'_{HK}$ &  \\
   & Thresh. 0.9, elimination Ppla, Phase $\log R'_{HK}$&  \\
FF' adapted from  & F' step on reference residuals & 5    \\
\hspace{0.5cm}  \cite{aigrain12} & FF' on original series, linear in $\log R'_{HK}$ &    \\
Binning  & Decreasing of the rotational modulation &    6 \\
GP & Rotation modulation, on original series  & 7    \\
GP & Rotation modulation, reference residuals & 7    \\
\hline
\end{tabular}
\end{center}
\tablefoot{Summary of the tested methods. All details are in the text. Only  the methods that are illustrated in Fig.~\ref{meth} have a number. 
}
\end{table*}

\subsection{Denoising of short-period peaks using $\log R'_{HK}$ periodograms}

\label{secA1}

We tested the denoising method for short periods on the residuals after correcting our reference model based on two series of peaks. First, we considered the list of peaks in the LSP of the $\log R'_{HK}$ time series, shorter than 50 days, and ordered by decreasing power. An iteration was then made on the peaks, starting with the highest and stopping at a given threshold. At each iteration, a sinusoidal fit was performed on the RV residuals at this period and was subtracted. We tested three thresholds: 1/ peaks higher than a level defined as 10\% of the highest peak, 2/ peaks higher than a noise level in the periodogram defined from the highest values in bins in the periodogram \footnote{50 bins are defined between 1 and 50 d, the maximum of the periodogram is computed in each bin, and the lowest of these 50 values is taken as a threshold.}, and 3/ peaks higher than the 1\% fap threshold computed on the LSP (computed with the bootstrap method assuming WGN used in Sect.\ref{sec4}).


The two other tests were performed based on the list of peaks in the LSP of the RV residuals computed from equation~\eqref{eqmodel}, also ordered by decreasing power. Two iterations  were performed: 1/ 
All peaks with  a period shorter than 50 days in the LSP and above the  threshold corresponding to an fap of 1\%  of the original RV time series were removed, and
2/ the same on the new residuals after first iteration (if performed).
The exact protocol may not be exactly similar to those used by other teams, but we tested several approaches, which should help to obtain robust results.


With the first and third thresholds based on the $\log R'_{HK}$ peaks, a very large number of peaks are removed (typically several hundreds), leading to a strong decrease in RV rms that is below our reference rms (provided by the correction performed  in Sect.~\ref{sec23}). However, the signal from the planet is also partially or totally  eliminated in the procedure, leading to a strongly underestimated mass estimate. On the other hand, with the other thresholds, only a small number of peaks are removed. The rms of the RV residuals are then decreased  by a few percent, but the impact on the mass is negligible and leads to no improvement of the final dispersion in mass. We conclude that these denoising methods are not efficient.

\subsection{Denoising based on correlations between RV and $\log R'_{HK}$ following the criterion of \cite{rosenthal21}}
\label{secA2}

\cite{rosenthal21} used a criterion based on the comparison between $\log R'_{HK}$ and RV time series to  attempt to reduce false positives in RV (they did not use it to correct the RV signal). If N peaks above the 1\% fap threshold in RV are observed, they applied the following criterion for each of these N peaks: They computed an RV model equal to the sum of the sinusoidal fits at each of the other (N-1) periods. This model was subtracted, and the correlation between the corresponding RV residuals and $\log R'_{HK}$ was computed. If it was high (i.e. close to 1, but the threshold was not specified), the method concludes  that the considered peak in RV might be due to stellar activity and  it was eliminated from the list of candidates.
In this section, we compute the same correlation for all the peaks above the 1\% fap threshold in the LSP  to verify whether this criterion can be used for denoising. If for a given peak the correlation as defined here is higher than a certain threshold, we subtract the sinusoidal fit at this period from the RV for denoising.

We applied this method to the original RV time series with  two thresholds: 1/ a constant threshold on the correlation of 0.9, and 2/ a variable threshold, chosen as the median on the N correlations, if there are N peaks above the 1\% fap threshold in the LSP, assuming that the planetary peak among the N peaks corresponds to one of the lowest correlations. This is justified below. 


The fixed threshold causes no  decrease in the rms compared to the original value. This is due to the fact that for most peaks above the fap threshold, the correlation is below the threshold, leading to no correction. As a consequence, the resulting fitted mass exhibits a very large dispersion without a peak at 4 M$_{\rm Earth}$ and an excess around zero.
There are indeed many peaks due to stellar activity. As a consequence, when considering a given peak due to activity, the model based on the other N-1 peaks is also mostly due to activity. The model based on one peak can therefore not be representative of the whole activity signal. It has hence a low correlation with the $\log R'_{HK}$ time series. 

We therefore studied the properties of the correlations for the different peaks of a given realisation in more detail. For each realisation, we first identified the correlation corresponding to the peak that was closest to the true planetary period and compared it to the other correlations. We found that the correlation for this peak is almost always the lowest one, and fewer than 2\% of the correlaations are above the median. We therefore implemented the second threshold, assuming that the selected peaks for denoising should not correspond to the peak due to the planet. This variable threshold allowed us to decrease the rms, but not as much as the reference method. In addition, the planetary mass distribution has a maximum around 0, meaning that the planetary signal is either removed in the residuals or remains hidden among the peaks due to activity. 

We conclude that this approach is  not satisfactory. We also conclude that although this method may be useful when the stellar signal is dominated by one period, for example the rotational modulation for young stars, a well-defined sinusoidal cycle, or with a poor temporal sampling leading to only one peak above the fap \cite[which may have been the case in][]{rosenthal21}, it is probably not efficient to remove false positives for a well-sampled solar-type star.

\subsection{Denoising based on correlations between RV and $\log R'_{HK}$ at specific periods}
\label{secA3}

Given the difficulty with the method described in Appendix~\ref{secA2} and the reason why it may not perform well, we attempted a variant of this method based on the comparison between peaks separately at each period. We considered here all RV peaks in the periodogram of the time series before correction. A sinusoidal fit was performed at each of the peaks, leading to an RV model for each period. A sinusoidal fit was then performed at each of these periods on the $\log R'_{HK}$ time series, leading to an activity model for each period. The correlation between the RV model and the $\log R'_{HK}$ model was computed for each period. If the correlation was high, the peak at this period was considered to be due to activity (this corresponds to the two models being in phase), and a denoising was performed at this period. We tested several variants of this approach: 1/ we  varied the threshold (0.8 and 0.9),  2/ we either considered all peaks or only peaks that were not in the true planet range (this last assumption would only be applicable for follow-ups as tested here, not for blind search), and  3/ the RV model for a given peak that was subtracted was either fitted on the RV residuals at this step (amplitude and phase) or was fitted with the phase derived from the $\log R'_{HK}$ fit. This led to eight variations.


None of these variations leads to good results. Even though the rms of the RV residuals is decreased, it is never as  small  as with the reference correction. Furthermore, the mass is not well estimated, with either an extremely wide peak or a peak maximum of the distribution around zero. We also checked the correlations (between the RV and $\log R'_{HK}$ models at a given period) for the peaks closest to the true planet period. They can take values very close to 1, so that the correlation (which is related to the phase between the two sinusoidal functions) is not a good criterion. This also shows that the use of the phase as a criterion to identify a false positive \cite[e.g. as in ][]{dumusque12} may not be  a robust criterion.

\subsection{Adapted FF' method}
\label{appC4}

\cite{aigrain12} proposed the FF' method, which basically works for simple activity pattern configurations, based on the following principle: F is the photometric signal that is  used to correct for the convective blueshift inhibition, and F' is the derivative of the photometric signal that is used to correct for the contrast contribution to RV measurements. Because the convective blueshift inhibition in plages is much better correlated to the $\log R'_{HK}$ than to the photometric signal \footnote{since the photometric signal is a residual, with a partial cancellation of the signal between spots and plages. }, we used the $\log R'_{HK}$  instead of F. The photometric signal was also produced in the simulations of Paper I and was therefore  used here to obtain F'. Because of noise and because the sampling is not regular, the computation of the derivative is noisy, however. We applied a spline interpolation to the time series, and in order to obtain the derivative at t, the average of two slopes (between t-1 day and t, and between t and t+1 day) was computed. The approximate computation of this derivative is therefore one limitation of the method.  

We applied this technique with two variations: 1/ The method was applied to the residuals computed previously based on equation~\eqref{eqmodel}, that is, we added a step by fitting only the model proportional to F'. 2/ the method was applied on the original RV time series, and the model was the sum of linear function of  $\log R'_{HK}$ and of a linear model in F' to avoid fitting too many parameters.  


We first consider the results  for $\Delta T_{\rm spot1}$ and the first method. In this case, the rms of the RV residuals is slightly decreased, by only a few percent, and the dispersion on the mass uncertainty is 5\%. The second approach decreases only more marginally, and there is no impact on the estimated masses. We also consider here $\Delta T_{\rm spot2}$ because we expect this method to be sensitive to the relative amplitude of the spot and plage contrasts to the convective blueshift inhibition contribution. This approach should therefore be more interesting for spots with high contrasts (or alternatively, for a star that is more spot dominated or has a low convective blueshift). In this case (highest spot contrast), both approaches lead to a decrease in the RV rms in the 10-13\% on average compared to the reference rms. The first approach performs best, with an improvement in the mass uncertainty by 9\%. The improvement is therefore only small.

\begin{figure}
\includegraphics{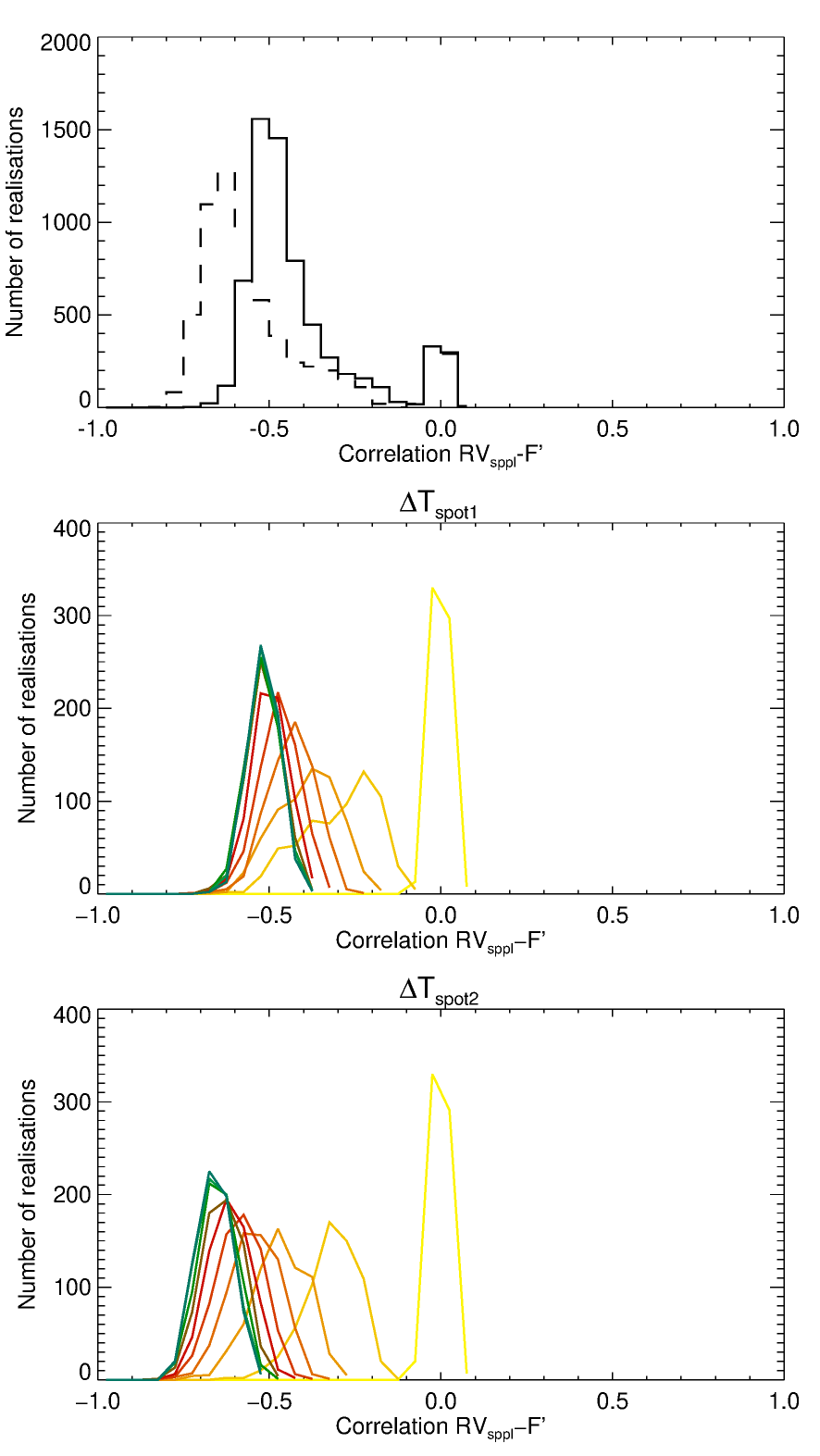}
\caption{
Distribution of the correlations between simulated RV$_{\rm sppl}$ and F' for G2 stars for all inclinations (upper panel, for $\Delta T_{\rm spot1}$ as a solid line and $\Delta T_{\rm spot2}$ as a dashed line)  and for $\Delta T_{\rm spot1}$ (second panel) and $\Delta T_{\rm spot2}$ (third panel) separately for the different inclinations: from pole-on (yellow) to edge-on (blue).
}
\label{sppl}
\end{figure}

\begin{figure}
\includegraphics{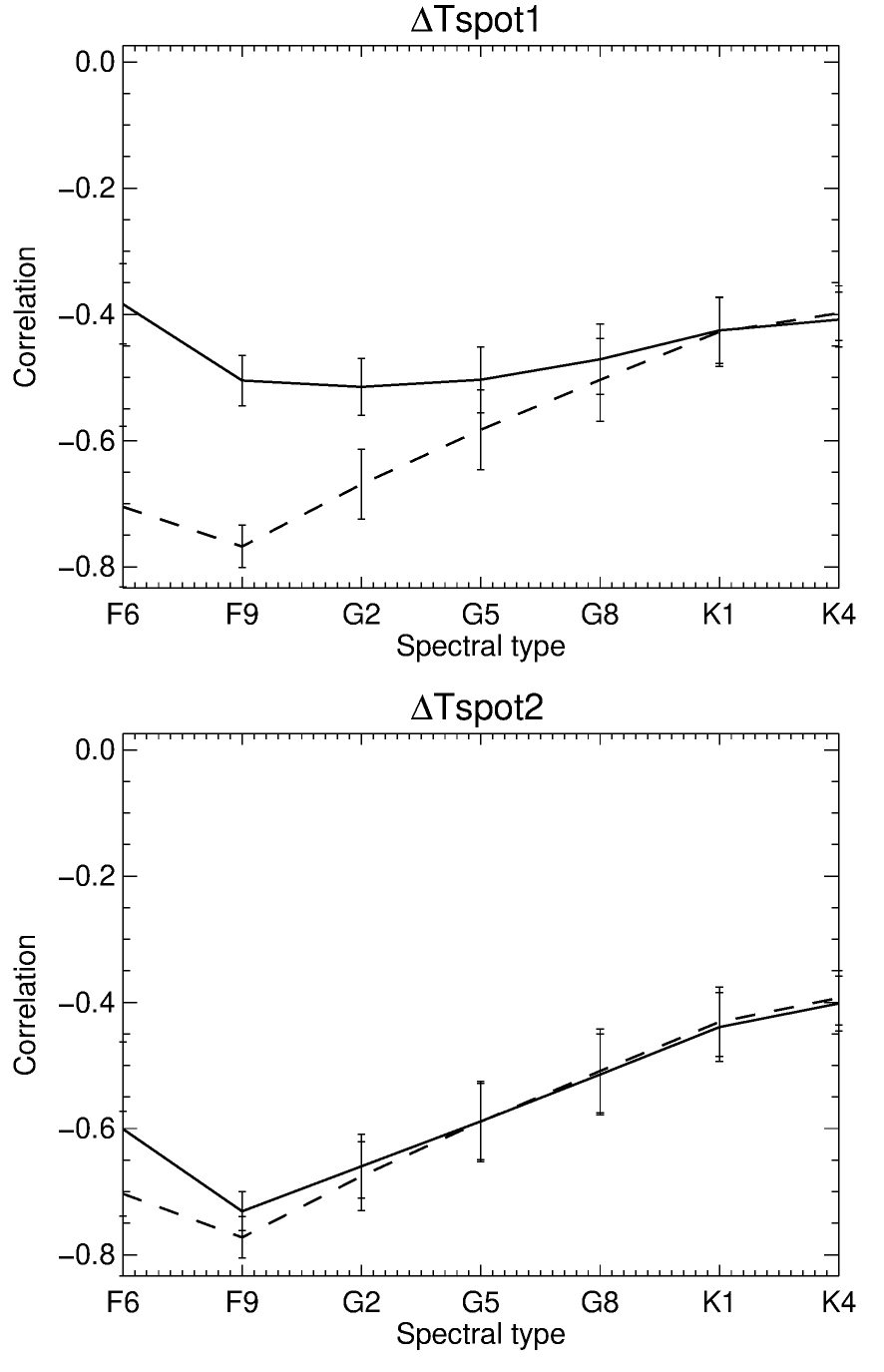}
\caption{
Average correlations between simulated RV$_{\rm sppl}$ (solid line) and F' vs. spectral type for $\Delta T_{\rm spot1}$ (upper panel) and $\Delta T_{\rm spot2}$ (lower panel). The dashed line shows only spots.  
}
\label{sppl2}
\end{figure}

To interpret this poor improvement and better understand  the limitations of this approach, we performed additional tests on our large set of simulations. We considered the time series with a  regular temporal sampling (with the original time step of one day on average, without a gap, so F' is less noisy). We computed F', and the contribution to RVs due to the contrast of spots and plages alone, hereafter RV$_{\rm sppl}$, and then the correlation between F' and RV$_{\rm sppl}$. The distributions of these correlations are shown in Fig.~\ref{sppl} for G2 stars, first for all inclinations, but separately for the two values of the spot contrast (the spots contribute more to the signal than the plage signal for $\Delta T_{\rm spot2}$), and then for different inclinations. Even if the signal is intrinsically anticorrelated in all cases 
\footnote{A dark spot leads to a decrease and then an increase in the photometric flux as it crosses the disk: F' is first negative and then positive. At the same time, the spot first blocks part of the disk coming towards us (leading to a redder spectrum and therefore a positive RV), and then blocks part of the disk going away from us (negative RV), hence the expected anticorrelation. }
(except as expected for the pole-on configuration), the anticorrelation is never excellent: for G2 stars and edge-on configurations, it is never below -0.8. The average correlation versus spectral type is shown in  Fig.~\ref{sppl2} (solid lines) and is compared to spots alone (dashed lines). We find that for spots alone, the best average correlations are close to -0.8 for F7 stars, but they are worse for other spectral types (and up to -0.4 for K4 stars). The anticorrelation was then degraded by the presence of plages, as shown by the difference between the two curves, and by the fact that the correlations are worse for $\Delta T_{\rm spot1}$, for which plages play a large role. These low correlations, obtained with a perfect sampling and no noise, explain why the method performs poorly, as we also expect these correlations to be degraded in the presence of noise and with a poorer sampling. This may be due to the presence of complex activity patterns, as discussed in \cite{aigrain12}. It could also be due to some fine effects in the difference in projection effects at the rotational timescale between RV on one side and $\log R'_{HK}$ or photometry on the other side, such as is observed for long-term variability between RV and $\log R'_{HK}$ \cite[][]{meunier19c}.


\subsection{Binning}
\label{appC5}

We averaged over the duration covering one rotation period (or more when the period was shorter than 15 days) to reduce both these effects and the remaining residuals due to rotation modulation. We typically obtained 6 to 11 observations per bin and a total of 90 to 180 bins. The rms of the RV residuals are naturally much lower than the reference rms due to the binning, but the uncertainty on the mass is similar, so that binning does not improve the results. The lower RV rms is 
compensated for by the lower number of points in the binned time series. 


\subsection{Gaussian processes}
\label{secA6}

Our objective here is not to be exhaustive because many variations exist, but first to test the performance of the correction of the variability of the rotation signal, as is usually done in the literature. We used the georges package \cite[][]{ambikasaran15} to train the GP on the $\log R'_{HK}$ time series, and then the RadVel package\footnote{\texttt{https://github.com/California-Planet-Search/radvel}} \cite[][]{fulton18} to simultaneously fit the GP and planetary signal on the RV time series. We note that this is very challenging because we analysed 1000-point time series, which would be extremely time consuming if we had used a number of chains (for the GP parameter optimisation by MCMC) as high as usual, even for a single time series, let alone for as many as 648. We therefore used a low number of walkers (typically twice the number of parameters), and a length of chains between 5000 and 8000. This may be different from an application on specific target in the literature, which usually has longer chains, but this was necessary in these first tests given the number of time series. We  performed various tests to confirm the convergence (longer chains on subset) to justify this compromise. In all tests below, the priors are usually considered to cover a very wide range. In all these tests, it is necessary to  jointly fit the parameters of the GP and of the planetary signal, otherwise, we found that this planetary signal was entirely absorbed by the GP. It is therefore not possible to use it to improve the detection rates because a reliable planetary signal must be identified prior to applying the method.  

Our first test was based on the $\log R'_{HK}$ and RV time series, again with a 4~M$_{\rm Earth}$ in the follow-up blind test (i.e. the planet has been detected, with a known period and phase). We focused on testing the impact on the long-period planet when we corrected for the rotational modulation, as is usually done in the literature. We followed the following protocol:
\begin{itemize}
\item{{\it Step 1}: The long-term variability is removed from the $\log R'_{HK}$ time series by fitting a two-sinusoidal function. The same function, scaled, is then subtracted from the RV time series.   }
 \item{{\it Step 2}: The hyperparameters describing a quasi-periodic kernel are adjusted with an MCMC minimisation method on the $\log R'_{HK}$. }
 \item{{\it Step 3}: The amplitude of the planetary signal is fitted on the RV residuals (obtained after step 1) to obtain a guess on K$_{\rm pla}$, the amplitude of  the planet signal. }
 \item{{\it Step 4}: The hyperparameters of the GP modelling stellar activity (also with a quasi-periodic kernel) and the parameters modelling the planetary signal (with K$_{\rm pla}$ as the only parameter) are then fitted to the RV residuals, from which the fitted mass can be determined, as well as the 1$\sigma$ uncertainties based on the corresponding quantiles from the MCMC results.  }
\end{itemize}

In the final step (step 4), we used the following quasi-periodic kernel of the RadVel package to model stellar activity:

\begin{equation} k(i,j) = \theta^{2}\cdot \exp\left(-\dfrac{|t_i-t_j|^{2}}{\lambda_e^{2}} - \dfrac{\sin\left(\dfrac{\pi|t_i-t_j|}{P_{rot}}\right)}{2\lambda_p^{2}}\right) + \gamma \cdot \delta_{i,j} ,\end{equation}

where the hyperparameters are the following: $\theta$ is the amplitude of the short-term (rotationally modulated) signal in m/s,
 $\lambda_e$ is the timescale of the evolution of the signal (in days), 
$\lambda_p$ is an dimensionless number representing the timescale of the variability during a rotation period, 
$P_{rot}$ is the periodicity of the signal (here the rotation period in days), and $\gamma$ is a Gaussian white noise in m/s, $\delta_{i,j} $ is the Kronecker delta function,
$t_i$ and $t_j$ are the time corresponding to each point in the time series, with i and j between 1 and 1000.
We refer to \cite{fulton18} for more details. For practical reasons,  step 2 was performed with another package, george\footnote{\texttt{https://github.com/dfm/george}} \cite[][]{ambikasaran15}, and therefore with a slightly different equation: it is also a quasi-periodic kernel, but with a change of the variables. For example it uses the log of the rotation period, and 1/2$\lambda_p^2$ is replaced by a parameter $\Gamma$. We do not detail them here because they are not studied in detail in the present paper.

The residuals after correction are much smaller than with the reference correction from Sect.~\ref{sec3}, that is, around 0.21 m/s on average. This is expected from the behaviour of the GPs, which are able to fit complex variability. This value is lower than what we would expect if the AR contribution were entirely corrected for, leaving the OGS and WGN signal in the residuals (which is about 0.35 m/s in those tests). However, the 1$\sigma$ uncertainty  on the mass (estimated from the distribution of the fitted masses as in Sect.~\ref{sec3}) is worse than for the reference analysis, it is about 46\% (1.8~M$_{\rm Earth}$), instead of 14\% for the reference analysis. This is most likely due to the fact that step 1 leaves some more long-term contribution. The average fitted mass is also worse, as illustrated in Fig.~\ref{meth} (vertical black line in the lower left panel).
We note that the individual uncertainties provided by the GP routine are comparable (the median is 1.7~M$_{\rm Earth}$ and the average 2.0~M$_{\rm Earth}$), showing that they are reliable. 

Because the long-term correction made in step 1 may have been too poor for the GP to be sufficient to provide a good performance, we then applied the same steps and kernel  to the residuals computed in Sect.~\ref{sec3} (equation~\ref{eqmodel}). 
We therefore followed step 2 (applied to the original $\log R'_{HK}$), and steps 3 and 4 (applied now to these RV residuals). The objective was to determine whether the uncertainty on the mass was improved with this additional step, allowing us to reduce the rms of the residuals, in particular, at short periods. The rms of the residuals  were then found to be low as well, around 0.30 m/s, that is, closer to what we expect for the OGS and WGN contributions. 
The 1$\sigma$ uncertainty on the mass (estimated from the distribution of the fitted masses as in Sect.~\ref{sec3}), 14\% (0.55~M$_{\rm Earth}$), was improved compared to the GP test including step 1, but is then very similar to the mass uncertainty obtained with the reference analysis (0.58~M$_{\rm Earth}$) despite the much lower rms of the residuals and a very good correction of the rotationally modulated stellar signal. The average of the estimated mass, 4.01~M$_{\rm Earth}$,  is improved compared to the reference case (3.8~M$_{\rm Earth}$), however. 
We note that the individual uncertainties provided by the GP routine are slightly larger (the median is 0.61~M$_{\rm Earth}$ and the average 0.82~M$_{\rm Earth}$).
The improvement  when applying this GP compared to the reference analysis is therefore only marginal. 
This is probably due to the fact that the residuals must also be improved at long timescales to provide a significant improvement. The possibility of exploring GP that work with both short (rotational) and long (cycle) timescales is beyond the scope of the present paper and will be explored in the future.

\section{Estimation of the cycle periods}

\label{secB}

\begin{figure}[h]
\includegraphics{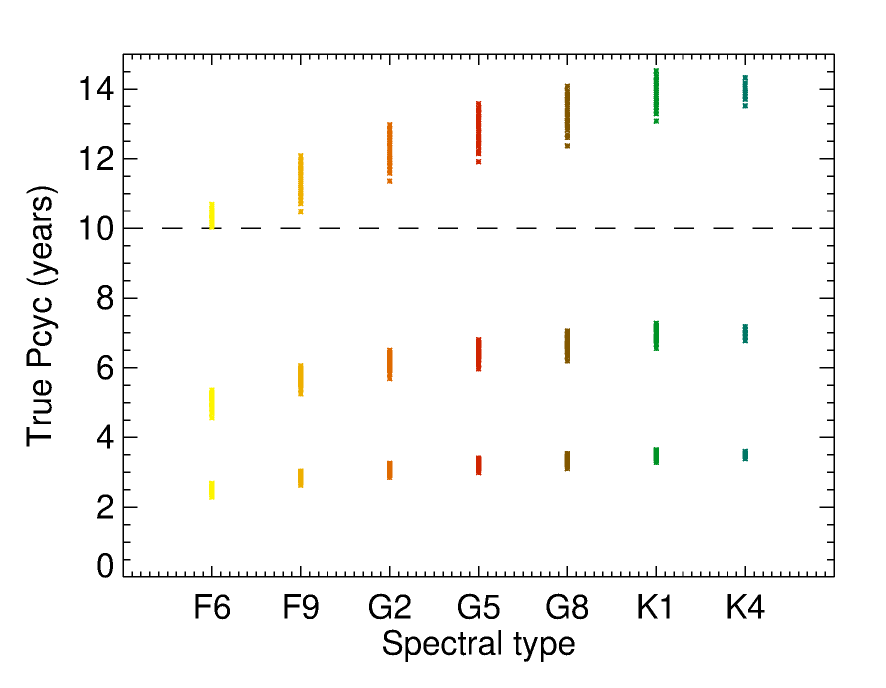}
\caption{
True cycle periods vs. spectral type  used in the set of simulations produced in Paper I. The dashed horizontal line is the time span used in the present paper.
}
\label{ptrue}
\end{figure}

\begin{figure}[h]
\includegraphics{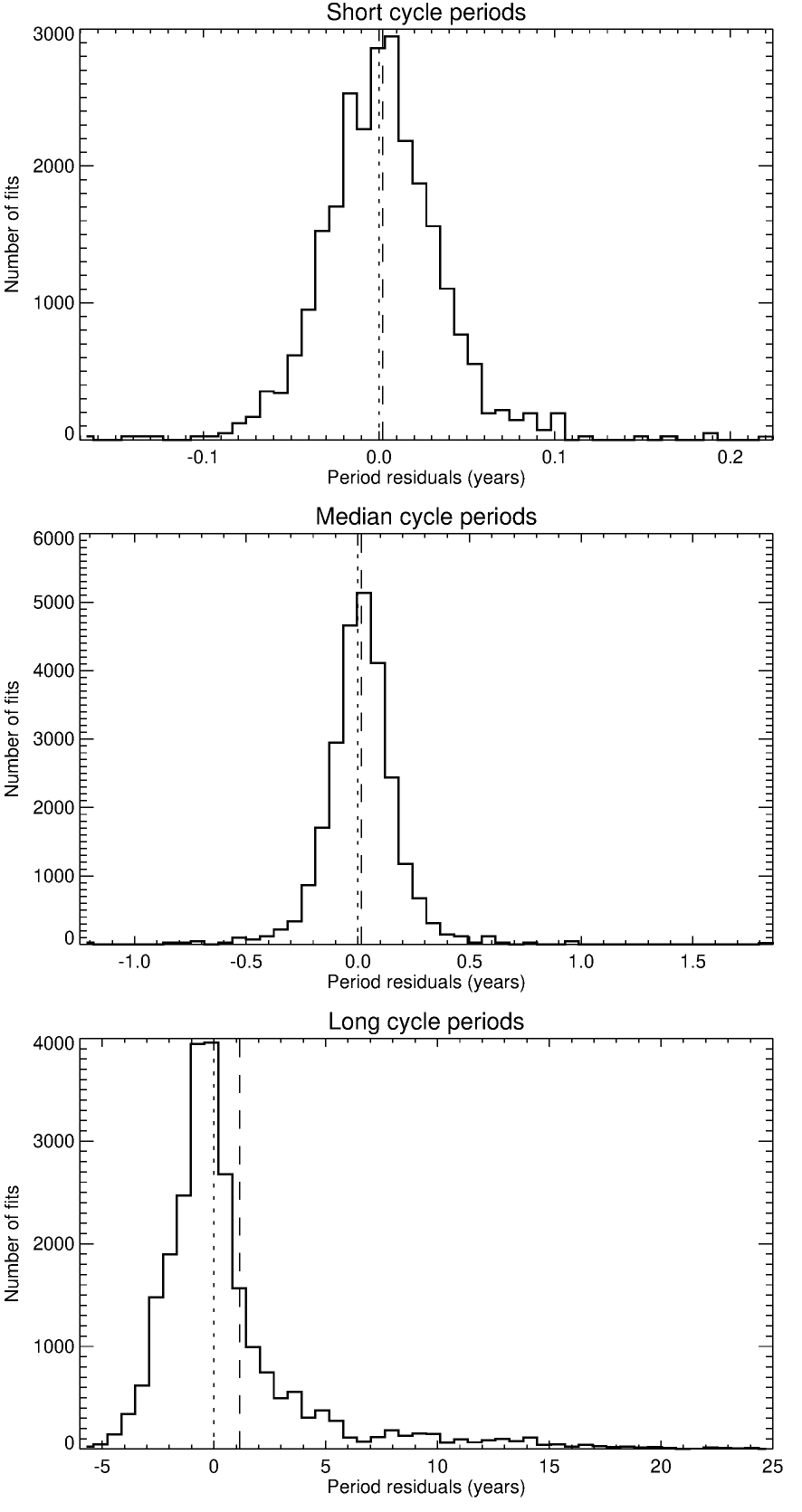}
\caption{
Distribution of the fitted minus true periods for the short periods (upper panel), median periods (middle panel), and long  periods (lower panel) for a long-term amplitude higher than 0.05. The plot is truncated for the last panel for clarity: about 0.7\% of the values form a long tail up to more than 200 years. The dashed vertical line is the mean value. 
}
\label{pfit_dist}
\end{figure}

\begin{table*}
\caption{Uncertainties on the cycle periods}
\label{tabcyc}
\begin{center}
\renewcommand{\footnoterule}{}  
\begin{tabular}{llccc}
\hline
Cycle  & Averaged & 1-$\sigma$ & 2-$\sigma$ & 3-$\sigma$  \\ 
   range &  P$_{\rm true}$ &  &  \\ \hline
 & &    \multicolumn{3}{c}{Negative $\sigma$} \\ \hline
Short & 3.1 & -0.03 (-0.03) & -0.11 (-0.07) & -1.92 (-0.14) \\
Median & 6.2 & -0.14 (-0.13) & -0.70 (-0.35) & -5.39 (-0.86) \\
Long & 12.6 & -1.73 (-1.59) & -6.43 (-3.25) & -12.20 (-4.78) \\
\hline
 & &    \multicolumn{3}{c}{Positive $\sigma$} \\ \hline
Short & 3.1 & +0.03 (+0.03)  & +0.18 (+0.08) & +8.57 (+0.19) \\
Median & 6.2 & +0.15 (+0.13) & +0.56 (+0.35) & +24.55 (+0.94) \\
Long & 12.6 & +3.00 (+2.93) & +16.66 (+14.3) & +123.03 (+87.3) \\
\hline
\end{tabular}
\end{center}
\tablefoot{Average true period and uncertainties (in years) for all spectral types, planet masses, and positions in the habitable zone. The total number of cycle estimates is 94824. The values in parentheses correspond to a selection of realisations with a long-term amplitude in $\log R'_{HK}$ higher than 0.05.
}
\end{table*}

Stellar cycle periods of solar-type stars are often derived from $\log R'_{HK}$ time series, either from very long surveys \cite[][]{baliunas95} or from surveys with a more limited duration \cite[e.g.][]{lovis11b,suarez18}. The correction method used in this work required estimating the cycle period from the $\log R'_{HK}$ time series in order to estimate the cycle phase for the model described in equation~\ref{eqmodel}. Furthermore, we recall that in our work, the time span of the synthetic time series is ten years, with 1000 nights of observations. The true cycle periods used to produce the synthetic time series are shown in Fig.~\ref{ptrue}. This is therefore a good opportunity to test the reliability of these estimates of the cycle periods on this large number of realisations for this type of stars. 

This was done as follows: The  $\log R'_{HK}$  time series were binned (30 days), and the LSP was computed. The period of the highest peak was used as a guess to a sinusoidal fit on the binned time series, from which a cycle period P$_{\rm fit}$ was estimated. Then we compared the values of P$_{\rm fit}$ obtained from the blind tests performed in Sect. 3 (i.e. for the star seen edge-on) with P$_{\rm true}$. 

We first considered all fits made for realisations corresponding to the seven spectral types, the four planet masses, and the three positions in the habitable zone (94824 periods). For the short and median periods P$_{\rm true}$, the fitted and true  periods agree in general. For the long  P$_{\rm true}$ values, which are all above the time span of the synthetic time series studied in this paper, although the estimation is good for a large fraction of the realisations, there are also many cases with a poor estimate, up to extremely high values above 200 years. As a consequence, there is only a very small bias for the short and median periods, but for the long periods (which are longer than our temporal coverage), the bias towards long periods is stronger (the average value differs by about one year from the true values). The distributions are shown in Fig.~\ref{pfit_dist} for the long-term amplitude, defined as the amplitude of the sinusoidal fit on the  $\log R'_{HK}$ time series (used as a proxy of the cycle amplitude) higher than 0.05. For the short and median periods, the distribution is close to a Gaussian, but it is very asymmetric for the high periods, with a long tail towards long periods and an excess of strong outliers. The performance is poorer for the small long-term amplitude time series, especially for the long periods. 

The 1$\sigma$, 2$\sigma$, and 3$\sigma$ levels corresponding to different selections to quantify the uncertainties are summarised in Table~\ref{tabcyc} for the whole sets of realisations and when selecting realisations with an amplitude higher than 0.05. For cycle periods of about 3 and 6 years, the 1$\sigma$ and 2$\sigma$ levels correspond to a good performance in terms of relative uncertainty. The 3$\sigma$ levels correspond to an uncertainty exceeding 100\%, however. For cycle periods of about 12 years, the 1$\sigma$ uncertainty is reasonable (smaller than 20\%), but the uncertainties are extremely large for the 2$\sigma$ and 3$\sigma$ levels.

\section{Additional residual analysis: Long-term residuals}
\label{appBB}

\begin{figure}
\includegraphics{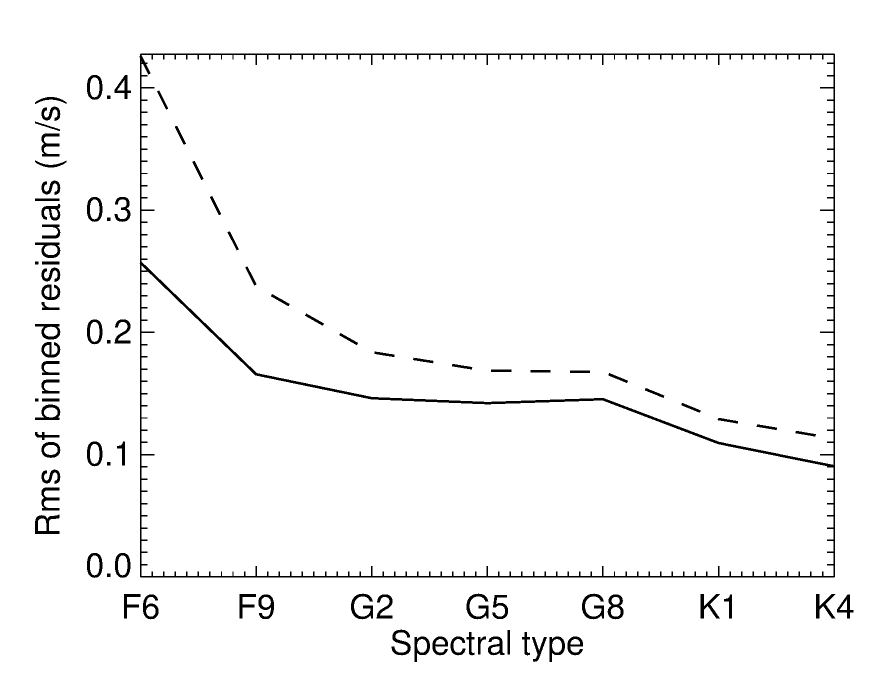}
\caption{
Rms of the binned RV residuals vs. spectral type from the follow-up analysis for $\Delta T_{\rm spot1}$ (solid line) and $\Delta T_{\rm spot2}$ (dashed line) for the middle of the habitable zone. The fitted 1~M$_{\rm Earth}$ planetary signal has been subtracted. }
\label{residusm}
\end{figure}

We present here an additional approach to analysing the residuals after corrections that is based on smoothed time series to focus on the long-term behaviour. 
We used the residuals from the follow-up blind tests, 1 M$_{\rm Earth}$ in the middle of the habitable zone (the other time series exhibit a similar behaviour). The planetary signal was also removed according to the fit.
 We first averaged the residuals into 150 d bins to eliminate the contribution of the remaining short-term variability. The chosen threshold of 150 d corresponds to at least three times the rotation period (and most of the times a larger factor), but is significantly shorter than the timescales involved in the long-term stellar variability. This threshold is therefore mostly pertinent for the longest planet periods among our input parameters. 
We then computed the RV rms on these smoothed time series. The objective was to estimate the order of magnitude of the amplitude of the long-term remaining signal and to compare it with the planetary signal. 
 Figure~\ref{residusm} shows the dependence of this long-term rms on spectral type for the two spot contrasts. 
 The rms decreases towards lower-mass stars. For G2 stars, it is at the level of 0.15 m/s (i.e. higher than the Earth signal).
 A similar computation performed on the residuals from the detection blind tests, not shown here, behaved similarly. The curves, which correspond to both spot contrasts, are close to the $\Delta T_{\rm spot1}$ curve in Fig.~\ref{residusm}, that is, they are lower than the average of the two curves because the average rms tends to be lower in these series (all inclinations instead of edge-on configurations).

\end{appendix}

\end{document}